\documentclass[a4paper,fleqn,usenatbib]{mnras}

\usepackage{newtxtext,newtxmath}

\usepackage[T1]{fontenc}
\usepackage{ae,aecompl}
\usepackage{hyperref}
\usepackage{multirow}

\usepackage{graphicx}	
\usepackage{amsmath}	
\usepackage{amssymb}	
\usepackage{multirow}
\usepackage{color}
\usepackage{natbib}

\usepackage[colorinlistoftodos]{todonotes}

\title[A catalogue of structural and morphological measurements for DES Y1]{A catalogue of structural and morphological measurements for DES Y1}

\author[DES Collaboration]{
	\parbox{\textwidth}{
		\Large
			\begin{center} F.~Tarsitano$^{1}$\thanks{federica.tarsitano@phys.ethz.ch},
		W.~G.~Hartley$^{2}$,\\
	\end{center}
		A.~Amara$^{1}$,
		A.~Bluck$^{3}$,
		C.~Bruderer$^{1}$,
		M.~Carollo$^{3}$,
		C.~Conselice$^{4}$,
		P.~Melchior$^{5}$,
		B.~Moraes$^{2}$,
		A.~Refregier$^{1}$,
		I.~Sevilla-Noarbe$^{6}$,
		J.~Woo$^{3}$,
		T.~M.~C.~Abbott$^{7}$,
		S.~Allam$^{8}$,
		J.~Annis$^{8}$,
		S.~Avila$^{9}$,
		M.~Banerji$^{10,11}$,
		E.~Bertin$^{12,13}$,
		D.~Brooks$^{2}$,
		D.~L.~Burke$^{14,15}$,
		A.~Carnero~Rosell$^{16,17}$,
		M.~Carrasco~Kind$^{23}$,
		J.~Carretero$^{18}$,
		C.~E.~Cunha$^{14}$,
		C.~B.~D'Andrea$^{19}$,
		L.~N.~da Costa$^{16,17}$,
		C.~Davis$^{14}$,
		J.~De~Vicente$^{6}$,
		S.~Desai$^{20}$,
		P.~Doel$^{2}$,
		J.~Estrada$^{8}$,
		J.~Frieman$^{8,21}$,
		J.~Garc\'ia-Bellido$^{22}$,
		D.~Gruen$^{14,15}$,
		R.~A.~Gruendl$^{23,24}$,
		G.~Gutierrez$^{8}$,
		D.~Hollowood$^{25}$,
		K.~Honscheid$^{26,27}$,
		D.~J.~James$^{28}$,
		T.~Jeltema$^{25}$,
		E.~Krause$^{29,30}$,
		K.~Kuehn$^{31}$,
		N.~Kuropatkin$^{8}$,
		O.~Lahav$^{2}$,
		M.~A.~G.~Maia$^{16,17}$,
		F.~Menanteau$^{23,24}$,
		R.~Miquel$^{32,18}$,
		A.~A.~Plazas$^{30}$,
		A.~K.~Romer$^{33}$,
		A.~Roodman$^{14,15}$,
		E.~Sanchez$^{6}$,
		B.~Santiago$^{34,16}$,
		R.~Schindler$^{15}$,
		M.~Smith$^{35}$,
		R.~C.~Smith$^{7}$,
		M.~Soares-Santos$^{36}$,
		F.~Sobreira$^{37,16}$,
		E.~Suchyta$^{38}$,
		M.~E.~C.~Swanson$^{24}$,
		G.~Tarle$^{39}$,
		D.~Thomas$^{9}$,
		V.~Vikram$^{40}$,
		A.~R.~Walker$^{7}$
		\begin{center} (DES Collaboration) \end{center}
	}
	\vspace{0.4cm}
	\\
	\parbox{\textwidth}{
		$^{1}$ Institute for Particle Physics and Astrophysics, ETH Zurich, Wolfgang-Pauli-Strasse 27, CH-8093 Zurich, Switzerland\\
		$^{2}$ Department of Physics \& Astronomy, University College London, Gower Street, London, WC1E 6BT, UK\\
		$^{3}$ Department of Physics, ETH Zurich, Wolfgang-Pauli-Strasse 16, CH-8093 Zurich, Switzerland\\
		$^{4}$ University of Nottingham, School of Physics and Astronomy, Nottingham NG7 2RD, UK\\
		$^{5}$ Department of Astrophysical Sciences, Princeton University, Peyton Hall, Princeton, NJ 08544, USA\\
		$^{6}$ Centro de Investigaciones Energ\'eticas, Medioambientales y Tecnol\'ogicas (CIEMAT), Madrid, Spain\\
		$^{7}$ Cerro Tololo Inter-American Observatory, National Optical Astronomy Observatory, Casilla 603, La Serena, Chile\\
		$^{8}$ Fermi National Accelerator Laboratory, P. O. Box 500, Batavia, IL 60510, USA\\
		$^{9}$ Institute of Cosmology \& Gravitation, University of Portsmouth, Portsmouth, PO1 3FX, UK\\
		$^{10}$ Institute of Astronomy, University of Cambridge, Madingley Road, Cambridge CB3 0HA, UK\\
		$^{11}$ Kavli Institute for Cosmology, University of Cambridge, Madingley Road, Cambridge CB3 0HA, UK\\
		$^{12}$ CNRS, UMR 7095, Institut d'Astrophysique de Paris, F-75014, Paris, France\\
		$^{13}$ Sorbonne Universit\'es, UPMC Univ Paris 06, UMR 7095, Institut d'Astrophysique de Paris, F-75014, Paris, France\\
		$^{14}$ Kavli Institute for Particle Astrophysics \& Cosmology, P. O. Box 2450, Stanford University, Stanford, CA 94305, USA\\
		$^{15}$ SLAC National Accelerator Laboratory, Menlo Park, CA 94025, USA\\
		$^{16}$ Laborat\'orio Interinstitucional de e-Astronomia - LIneA, Rua Gal. Jos\'e Cristino 77, Rio de Janeiro, RJ - 20921-400, Brazil\\
		$^{17}$ Observat\'orio Nacional, Rua Gal. Jos\'e Cristino 77, Rio de Janeiro, RJ - 20921-400, Brazil\\
		$^{18}$ Institut de F\'{\i}sica d'Altes Energies (IFAE), The Barcelona Institute of Science and Technology, Campus UAB, 08193 Bellaterra (Barcelona) Spain\\
		$^{19}$ Department of Physics and Astronomy, University of Pennsylvania, Philadelphia, PA 19104, USA\\
		$^{20}$ Department of Physics, IIT Hyderabad, Kandi, Telangana 502285, India\\
		$^{21}$ Kavli Institute for Cosmological Physics, University of Chicago, Chicago, IL 60637, USA\\
		$^{22}$ Instituto de Fisica Teorica UAM/CSIC, Universidad Autonoma de Madrid, 28049 Madrid, Spain\\
		$^{23}$ Department of Astronomy, University of Illinois at Urbana-Champaign, 1002 W. Green Street, Urbana, IL 61801, USA\\
		$^{24}$ National Center for Supercomputing Applications, 1205 West Clark St., Urbana, IL 61801, USA\\
		$^{25}$ Santa Cruz Institute for Particle Physics, Santa Cruz, CA 95064, USA\\
		$^{26}$ Center for Cosmology and Astro-Particle Physics, The Ohio State University, Columbus, OH 43210, USA\\
		$^{27}$ Department of Physics, The Ohio State University, Columbus, OH 43210, USA\\
		$^{28}$ Harvard-Smithsonian Center for Astrophysics, Cambridge, MA 02138, USA\\
		$^{29}$ Department of Astronomy/Steward Observatory, 933 North Cherry Avenue, Tucson, AZ 85721-0065, USA\\
		$^{30}$ Jet Propulsion Laboratory, California Institute of Technology, 4800 Oak Grove Dr., Pasadena, CA 91109, USA\\
		$^{31}$ Australian Astronomical Observatory, North Ryde, NSW 2113, Australia\\
		$^{32}$ Instituci\'o Catalana de Recerca i Estudis Avan\c{c}ats, E-08010 Barcelona, Spain\\
		$^{33}$ Department of Physics and Astronomy, Pevensey Building, University of Sussex, Brighton, BN1 9QH, UK\\
		$^{34}$ Instituto de F\'\i sica, UFRGS, Caixa Postal 15051, Porto Alegre, RS - 91501-970, Brazil\\
		$^{35}$ School of Physics and Astronomy, University of Southampton,  Southampton, SO17 1BJ, UK\\
		$^{36}$ Brandeis University, Physics Department, 415 South Street, Waltham MA 02453\\
		$^{37}$ Instituto de F\'isica Gleb Wataghin, Universidade Estadual de Campinas, 13083-859, Campinas, SP, Brazil\\
		$^{38}$ Computer Science and Mathematics Division, Oak Ridge National Laboratory, Oak Ridge, TN 37831\\
		$^{39}$ Department of Physics, University of Michigan, Ann Arbor, MI 48109, USA\\
		$^{40}$ Argonne National Laboratory, 9700 South Cass Avenue, Lemont, IL 60439, USA\\
	}
}

\date{Accepted XXX. Received YYY; in original form ZZZ}

\pubyear{2018}

\begin{document}
\label{firstpage}
\pagerange{\pageref{firstpage}--\pageref{lastpage}}
\maketitle

\begin{abstract}
	We present a structural and morphological catalogue for 45 million objects selected from the first year of data from the Dark Energy Survey (DES). Single S\'ersic fits and non-parametric measurements are produced for \textit{g}, \textit{r} and \textit{i} filters. The parameters from the best-fitting S\'ersic model (total magnitude, half-light radius, S\'ersic index, axis ratio and position angle) are measured with \textsc{Galfit}; the non-parametric coefficients (concentration, asymmetry, clumpiness, Gini, M20) are provided using the Zurich Estimator of Structural Types (ZEST+). To study the statistical uncertainties, we consider a sample of state-of-the-art image simulations with a realistic distribution in the input parameter space and then process and analyse them as we do with real data: this enables us to quantify the observational biases due to PSF blurring  and magnitude effects and correct the measurements as a function of magnitude, galaxy size, S\'ersic index (concentration for the analysis of the non-parametric measurements) and ellipticity.
	We present the largest structural catalogue to date: we find that accurate and complete measurements for all the structural parameters are typically obtained for galaxies with \textsc{SExtractor} $\texttt{MAG\_AUTO\_I} \le 21$. Indeed, the parameters in the filters \textit{i} and \textit{r} can be overall well recovered up to $\texttt{MAG\_AUTO} \le 21.5$, corresponding to a fitting completeness of $\sim 90\%$ below this threshold, for a total of 25 million galaxies. The combination of parametric and non-parametric structural measurements makes this catalogue an important instrument to explore and understand how galaxies form and evolve. The catalogue described in this paper will be publicly released alongside the Dark Energy Survey collaboration Y1 cosmology data products at the following URL: \url{https://des.ncsa.illinois.edu/releases/y1a1/gold/morphology}.
\end{abstract}




\section{Introduction}
Any explanation of the formation and evolution of galaxies must necessarily include a description of the diverse forms that galaxies take. The morphology of the luminous components of a galaxy, including its classification or decomposition into a bulge and disk (e.g., \citealp{Kormendy1977, deJong}) or identification of features such as bars, rings or lenses (e.g., \citealp{Kormendy1979, Combes1981, Elmegreen1996}), are a result of its aggregated formation history. Assigning meaningful morphological types or quantifying the distribution of light across the extent of a population of galaxies, is therefore of fundamental importance in understanding the processes that govern their evolution. 

A quantitative description of galaxy morphology is typically expressed in terms of structural parameters (brightness, size, shape) and properties of the light distribution (concentration, asymmetry and clumpiness), though human classifications are still used. The development of \textit{citizen science} projects like Galaxy Zoo \citep{LintottKevin, Simmons, Willett} and sophisticated machine learning algorithms \citep{Lahav1,Lahav2,HC1,HC2,Manda2010,Dieleman} have helped to maintain the relevance of these perception-based morphologies in the current literature. Nevertheless, most recent work on the subject of galaxy morphologies rely on either \textit{parametric} or \textit{non-parametric} approaches to quantify the galaxy's light distribution.

Parametric methods fit two-dimensional analytic functions to galaxy images. The mathematical model of the light fall-off is convolved with the point spread function (PSF) to take into account the seeing.
The most general assumed function for this purpose is the S\'ersic profile \citep{Sersic}. The second class, non-parametric methods, perform an analysis of the light distribution within a certain elliptical area, usually defined through the Petrosian radius associated with the galaxy. Common estimates are of the degree to which the light is concentrated, quantifying the asymmetry of the light distribution and searching for clumpy regions: this method is called \emph{CAS system} (Concentration, Asymmetry and Smoothness or Clumpiness) and can be extended with further parameters, Gini and M20 \citep{Conselice2003, Abraham2003, Lotz2004, Law2007}. These parameters together can describe the major features of galaxy structure without resorting to model assumptions about the galaxy's underlying form, as is done with the S\'ersic profile. However, they are determined without a PSF deconvolution and need an additional calibration.

Even alone, distributions of morphological quantities represent powerful constraints on possible galaxy formation scenarios. But combined with other physical quantities, they can provide key insights into the processes at play, supporting or even opening new ideas on evolutionary mechanisms (\citealp{Kauffmann2004,Weinmann2006,Kevin2007,vanderWel2008p2,vanderWel2008,Bamford,Kevin2014}). For instance, the relationship between the masses, luminosities and sizes of massive disks and spheroids suggests dissipative formation processes within hierarchical dark matter assembly \citep{WhiteRees1978,FallEfstathiou1980} or the occurrence of galaxy-galaxy mergers \citep{Toomre1972,Toomre, Barnes,NaabBurkert,Conselice2003,Lin2004,Conselice2008,Conselice2008Two,Jogee2009,Jogee2009p2}. On the other hand, analysing galaxy sub-structure (e.g. with a bulge + disk decomposition) can open up evidence of further mechanisms: bulges, disks and bars may be formed by secular evolution processes \citep{Kormendy1979, Kormendy2004, Bournaud, Genzel, Fisher, Sellwood} or by the interplay between smooth and clumpy cold streams and disk instabilities \citep{Dekel2009,Dekel2-2009}. In this sense bulges may be formed without major galaxy mergers, as is often thought.

Of particular interest in recent years, have been the questions over the degree to which galaxy environment impacts upon morphology \citep[e.g.][]{Dressler80, Postman05, Lani13, Kuutma17}, and the connection between morphology and cessation of star formation in galaxies \citep[e.g.][]{Blanton03, Martig09, Bell12, Woo15}. Faced with often subtle correlations or hidden variables within strong correlations, these questions demand far greater statistical power and measurement precision than had been possible from the available data sets in the preceding decades. These demands require efficient pipelines to automate and streamline the analysis of large astronomical data sets. GALAPAGOS \citep{Gray2009,Boris2011,Barden2012} is perhaps the most widely used of such pipelines. It offers a routine to simplify the process of source detection, to cut postage stamps, prepare masks for neighbours if needed and estimate a robust sky background and has been used at both low redshift in the GEMS survey \citep{Boris2007}, and at higher redshift on the CANDELS \citep{vanderWel12} data. 

At low redshift the state-of-the-art to date are the catalogues constructed from Sloan Digital Sky Survey (SDSS, \citealt{York00}) data, in particular the bulge+disk catalogue of \cite{Simard2011} numbering almost 1 million galaxies. Such statistical power has been lacking at higher redshifts, but the advent of large-scale cosmology experiments optimised for weak lensing analyses, such as the Dark Energy Survey (DES) and Hyper Suprime-Cam (HSC) \citep{HSC2012}, provide a great opportunity to fill in much of this gap. DES is the largest galaxy survey to date, with a narrower PSF and images typically two magnitudes deeper than the SDSS.

In order to create as complete a set of structural measurements for DES as possible we adopt both parametric and non-parametric approaches, using the software \textsc{Galfit} \citep{Peng124,Peng2010} for parametric S\'ersic fitting and \textsc{ZEST+} for a non-parametric analysis of the structural properties of our galaxy sample. The first provides us with the measurements of the magnitude, effective radius, S\'ersic Index, axis ratio and orientation angle of the galaxy; the second one outputs an extended set of parameters, completing the CAS system with Gini and M20, plus the values of magnitude, half light radius and ellipticity, measured within the galaxy Petrosian ellipse.

The scale of the DES data set requires a new dedicated pipeline in order to handle the DES data structure, optimise run-time performance, automate the process of identifying and handling neighbouring sources and prepare tailored postage stamps for input to the two software packages. The resulting dataset is by far the largest catalogue of structural parameters measured to date, numbering 45 million galaxies, which exceeds previous catalogues by more than an order of magnitude in size, and reaches redshift, $z\sim1$. It includes parametric and non-parametric measurements in three photometric bands, intended to be used in concert and to provide a comprehensive view of the galaxies' morphologies. In this sense, our catalogue constitutes a significant step in our capabilities to study the nature of galaxy morphology in the Universe.

This paper is structured as follows: in Section~\ref{sec:Data} we give an overview of the Dark Energy Survey, describing the data and the image simulation data we used for this work. In Section~\ref{sec:datapreparation} we focus on the details of our sample selection and pre-fitting routine, presenting the algorithms developed to prepare and process the data. Sections~\ref{sec:tools} and~\ref{sec:nonparams} are dedicated to the parametric and non-parametric fits, respectively.
In each of these two sections, we present a detailed description of the fitting software used for this work, discuss the completeness and validation of the fitted sample from each method, provide an overview of the characteristics of the catalogue and perform a calibration of the output quantities with image simulations. The calibration for the \textit{i} band are shown in those sections; Appendix~\ref{sec:calibrations_gr} includes the calibration maps also for the \textit{g} and \textit{r} filters.  Section~\ref{sec:nonparams} also introduces a set of basic cuts as a starting point in building a science-ready sample. Finally in Section~\ref{sec:conclusions} we summarise our work. A manual explaining the catalogue columns is presented in Appendix~\ref{sec:cataloguemanual}.

\section{Data}
\label{sec:Data}
\subsection{The Dark Energy Survey}

The Dark Energy Survey (DES) \citep{Diehl2005,Abbott2016} is a wide-field optical imaging survey covering $5000$ deg$^{2}$ of the southern equatorial hemisphere in \textit{grizY} bands\footnote{\href{http://www.darkenergysurvey.org}{http://www.darkenergysurvey.org}}. Survey observations began in August 2013 and over five years it will provide images of 300 million of galaxies up to redshift $\sim1.4$ \citep{Diehl2014}.
The survey is designed to have a combination of area, depth and image quality optimized for cosmology, and in particular the measurement of weak gravitational lensing shear. However, its rich data set is well-suited to many areas of astronomy, including galaxy evolution, Milky Way and Local Group science, stellar populations and Solar System science  \citep{Abbott}.
\\
DES uses the Dark Energy Camera (DECam), a mosaic imager with a $2.2^{\circ}$ diameter field of view and a pixel scale of $0.263''$ per pixel mounted at the prime focus of the Victor M. Blanco 4m Telescope at Cerro Tololo Inter-American Observatory. During the requested 525 observing nights it is expected to reach photometric limits of $g=24.6$, $r=24.4$, $i=23.7$, $z=22.7$ and $Y=21.5$ ($10\sigma$ limits in $1.5''$ apertures assuming $0.9''$ seeing) following ten single-epoch exposures of 90 seconds each for $griz$ and 45 seconds each for $Y$ \citep{Flaugher}.

The DES data are processed, calibrated and archived through the DES Data Management (DESDM) system \citep{Drlica-Wagner, Morganson2018}, consisting of an image processing pipeline which performs image de-trending, astrometric calibration, photometric calibration, image co-addition and \textsc{SExtractor} \citep{Bertin} catalogue creation. 
The DESDM imaging co-addition combines overlapping single-epoch images in a given filter, which are then remapped to artificial tiles in the sky so that one co-add image per band is produced for every tile. These tiles are padded to ensure that each object is entirely contained in at least one tile, but also results in a small fraction of duplicate objects found in different tiles which are removed at a later stage. In order to account for PSF variations caused by object location in the focal plane and the combination of images with different seeing, the catalogue creation process uses \texttt{PSFEx} \citep{Bertin2011, Bertin2013} to model the PSF. \texttt{PSFex} produces a basis set of model components on the same pixel scale as the science image that are combined via linear combination into a location-dependent PSF. The final step combines the photometry of each co-add object into a single entry in multi-wavelength \textsc{SExtractor} catalogues. For more details about the DESDM co-addition and PSF modelling we refer the reader to \citealt{Sevilla:2011ps}, \citealt{Desai} and \citealt{Mohr}. \\

In this work we use the \texttt{DES Y1A1 COADD OBJECTS} data release, comprising 139,142,161 unique objects spread over about $1800 \ deg^{2}$  in 3707 co-add tiles, constructed from the first year of DES survey operations. The tiles are combinations of 1-5 exposures in each of the \textit{grizY} filters and the average coverage depth at each point in the retained footprint is $\sim3.5$ exposures. We consider 3690 tiles in total: the catalogue for the remaining tiles, located in the $30~deg^{2}$ of cadenced supernovae fields, will be presented in future work. The data include all the products of the DESDM pipeline and imaging co-addition (the co-add tiles and their respective segmentation maps, the PSF models and the \textsc{SExtractor} catalogues), plus the \texttt{Y1A1 GOLD} catalogue (\citealt{Drlica-Wagner}). In the  \texttt{Y1A1 GOLD} catalogue, the data collected in DES year-one have been characterised and calibrated in order to form a sample which minimises the occurrence of artefacts and systematic features in the images. It further provides value-added quantities such as the star-galaxy classifier \texttt{MODEST} and photo-z estimates. \texttt{GOLD} magnitudes are corrected for interstellar extinction using stellar locus regression (SLR) \citep{High2009}. We combine the \textsc{SExtractor} DESDM catalogues with the \texttt{Y1A1 GOLD} catalogue to make the sample selection, as described in section~\ref{subsec:sampleselection}, and we also benefit from the application of the \texttt{MODEST} classifier during the analysis of the completeness of our fitting results, reported in more detail in section~\ref{sec:parametric_completeness}.

\subsection{Image simulation data}
\label{sec:simulations}
In fitting galaxy light profiles, faint magnitude regimes are well known to present larger systematic errors in the recovered galaxy sizes, fluxes and ellipticities \citep{Bernstein2002, Boris2007, MelchiorViola2012}. A larger FWHM of the PSF can also introduce increased uncertainties and systematic errors during morphological estimation. In order to overcome these issues we use sophisticated image simulations to derive multi-parameter vectors that quantify any biases arising from our analyses, data quality or modelling assumptions. The simulations we use for this purpose are produced by the \textit{Ultra Fast Image Generator} (\textsc{UFig}) \citep{Berge2013} run on the Blind Cosmology Challenge simulation (\textsc{BCC}, \citealt{Busha2013}) and released for DES Y1 as \textsc{UFig-BCC}.\\
\textsc{UFig-BCC} covers an area of $1750 \ deg^{2}$ and includes images which are calibrated to match the DES Y1 instrumental effects, galaxy distribution and survey characteristics. Briefly, an input catalogue of galaxies is generated based on the results of an N-body simulation with an algorithm to reproduce the observed luminosity and colour-density relations.

\section{Pre-fitting Pipeline}
\label{sec:datapreparation}
In this section we describe first the sample selection we apply to the \texttt{DES Y1A1 COADD OBJECTS}, discussing the cuts applied and the initial distributions. Then we describe the process which prepares the data to be fitted both with parametric and non-parametric approach.

\subsection{Sample Selection}
\label{subsec:sampleselection}

\begin{figure*}
	\includegraphics[width=\textwidth]{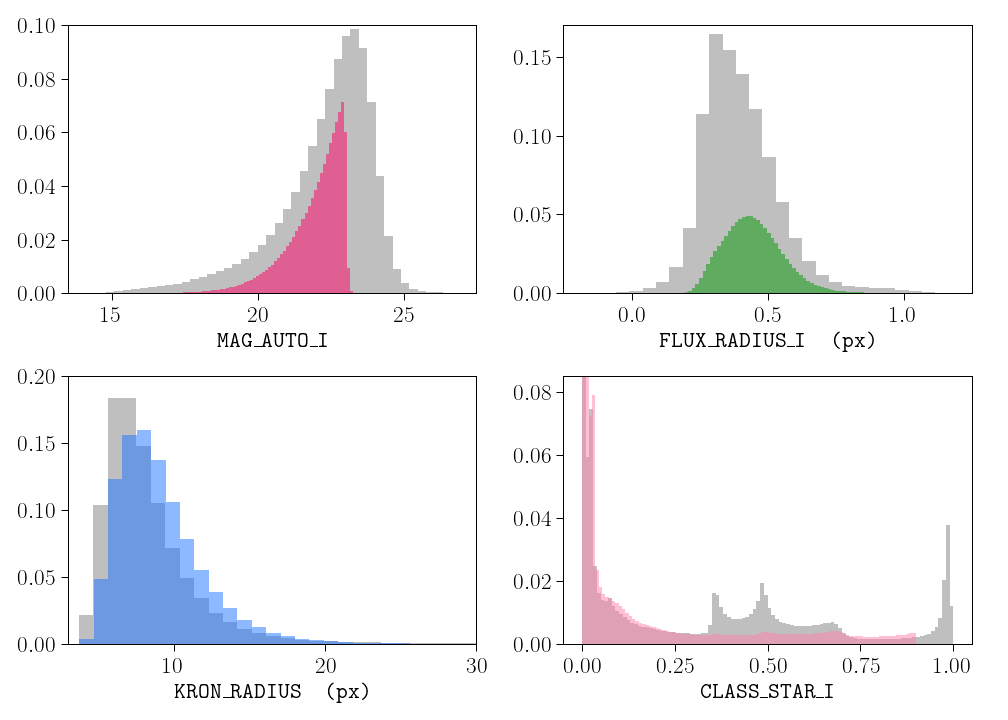}
	\caption{Normalised distributions of the variables involved in the sample selection in the \textit{i} band. From upper left to bottom right: \texttt{MAG\_AUTO}, \texttt{CLASS\_STAR}, \texttt{FLUX\_RADIUS} and \texttt{KRON\_RADIUS}. The cuts applied to each variable are described in more detail in section~\ref{subsec:sampleselection} and summarised in Table~\ref{table:sampleselection}. In each panel the grey histogram refers to the whole dataset, while the coloured one represents the distribution in that variable for the selected sample.}
	\label{fig:selhistos}
\end{figure*}

\begin{table}
	\begin{center}
		\begin{tabular}{| l | l | l | l |}
			\hline
			\hline
			SELECTION TYPE & \color{black} SELECTION CUT  \\ \hline
			Gold match  & \texttt{IN\_GOLD} = True   \\ \hline
			Image flags & \texttt{FLAGS\_x} = 0  \\ \hline
			S/G & \texttt{CLASS\_STAR\_i} $\le 0.9$  \\ \hline
			Magnitude  & \texttt{MAG\_AUTO\_i} $\le 23$  \\ \hline
			Size (I) & \texttt{FLUX\_RADIUS} > 0 px   \\ \hline
			Size (II) & \texttt{KRON\_RADIUS} > 0 px   \\ \hline
			Regions & \texttt{FLAGS\_BADREGION} = 0  \\ \hline
			\hline
		\end{tabular}
		\caption{Summary of the cuts applied to the overlapping sample between the catalogue provided by the DESDM pipeline and the \texttt{Y1A1 GOLD} catalogue. The selected objects must satisfy the requirements described in section~\ref{subsec:sampleselection}. \textit{x} identifies the filter ($x=g,r,i$).}
		\label{table:sampleselection}
	\end{center}
\end{table}
For this work we use a tile-by-tile approach, independently for each filter: every step from the sample selection itself to the fitting process is performed separately in each tile and band, with the exception of an overall \textit{i}-band magnitude cut and fiducial star-galaxy separation. We organise the \texttt{Y1A1 GOLD} catalogue into sub-catalogues to include the objects in each co-add tile and match them with the relevant DESDM \textsc{SExtractor} catalogues, extracted from that tile. We apply cuts to specific flags in the catalogues and to the parameters we use as priors for the fits in order to remove the most probable point-like sources, whilst avoiding removing galaxies. In addition we remove a small amount of the survey area in order to work with objects whose \textsc{SExtractor} detection and images are reliable and well-suited for the fitting process.
An object is selected if it fulfils the following requirements:
\begin{itemize}
	\item $\texttt{FLAGS\_X} = 0$;
	\item $\texttt{GOLD\_MAG\_AUTO\_I} \le 23$;
	\item $\texttt{FLUX\_RADIUS\_X} > 0$;
	\item $\texttt{KRON\_RADIUS\_X} > 0$;
	\item $\texttt{CLASS\_STAR\_I} < 0.9$;
	\item $\texttt{FLAGS\_BADREGION} = 0$,	
\end{itemize}
where $X=g,r,i,z,Y$. The cut in \texttt{FLAGS} removes objects that are either saturated, truncated or have been de-blended.
We apply the cuts using the \textit{i} band as our reference band; indeed the seeing FWHM in this filter is on average the smallest of the five bands. In using the \texttt{CLASS\_STAR} classifier at this stage we perform a conservative star-galaxy discrimination (S/G), so that we attempt a fit for any object which could be a galaxy. During the validation analysis we will remove further objects, applying a stricter classifier, named \texttt{MODEST}.
We refer to section~\ref{sec:stellarcontamination} for its definition and more details about its impact on this work.
By \texttt{GOLD\_MAG\_AUTO} we refer to the \textsc{SExtractor} quantity \texttt{MAG\_AUTO}, corrected by photometric calibration through SLR as provided by the \texttt{Y1A1 GOLD} catalogue (\citealt{Drlica-Wagner}). In the following sections we will simply use the original uncalibrated \textsc{SExtractor} \texttt{MAG\_AUTO}. The \texttt{AUTO} photometry is calculated with an elliptical aperture of radius, 2.5 Kron radii. \texttt{FLUX\_RADIUS} is the circular radius that encloses half of the light within in the AUTO aperture. Throughout this work, we use \texttt{KRON\_RADIUS} to refer to the semi-major axis of the Kron ellipse, i.e. the \textsc{SExtractor} values \texttt{A\_IMAGE} and \texttt{KRON\_RADIUS} multiplied together.\\
\texttt{FLAGS\_BADREGION} is a flag from the \texttt{Y1A1 GOLD} catalogue tracing the objects that lie in problematic areas, which are close to high-density stellar regions and/or present ghosts and glints.
The sample selection cuts described above are summarised in Table~\ref{table:sampleselection}. The normalised distributions of the variables considered during the initial cuts, comparing the selected sample of 45 million objects with the entire dataset (in grey), are shown in Fig.~\ref{fig:selhistos}. 

\subsection{Data processing}
\label{sec:prefittingpip}
The co-add data used in this work are processed in a dedicated pre-fitting pipeline, called Selection And Neighbours Detection (SAND), which has been developed in order to prepare the postage stamps to be fit, their ancillary files in the formats required by \textsc{Galfit} and \textsc{ZEST+} and perform essential book-keeping operations. The pipeline performs three steps: \textit{sample selection} (as described in section~\ref{subsec:sampleselection}), \textit{stamp cutting} and \textit{identification of neighbouring sources}. It is important to note that the objects excluded by our initial sample selection (section \ref{subsec:sampleselection}) are still fit as neighbouring objects where appropriate. For this reason dedicated flags are assigned to each object in the sample, in order to trace their \texttt{CLASS\_STAR} classification and possible anomalies in their photometric and structural properties. Collectively, we refer to these flags as $\texttt{STATUS\_FLAGS}$, and document the components and possible values in Appendix~\ref{sec:cataloguemanual}. 

For each selected object, an image postage stamp is created, initially with half-width equal to 3 times its Kron radius\footnote{i.e. \textsc{SExtractor} $\texttt{KRON\_RADIUS} \times \texttt{A\_IMAGE}$.}. Using the relevant segmentation map, the algorithm calculates the percentage of pixels that are not associated with sources (i.e. are background pixels) and approves the stamp if the sky fraction is at least $60\%$. Otherwise, the image stamp is rejected and is enlarged in size in integer multiples of Kron radius until this requirement is satisfied.

The last step of the pre-fitting routine is dedicated to the identification and cataloguing of neighbours: using the postage segmentation maps it locates the neighbouring objects and, with the above mentioned \texttt{STATUS\_FLAGS}, identifies nearby potential stars and/or galaxies with unreliable \textsc{SExtractor} detection. With this last expression we refer to the objects which have unphysical \textsc{SExtractor} parameters (negative sizes, magnitude set to standard error values) and/or are flagged as truncated or saturated objects. In addition to their coordinates and \textsc{SExtractor} properties, the routine catalogues the relative \textsc{SExtractor} magnitude and the presence of overlapping Kron-like isophotes between the central galaxy and its neighbours: these cases are then classified with two dedicated flags, called \texttt{ELLIPSE\_FLAGS} and \texttt{MAX\_OVERLAP\_PERC}, which are fully described in Appendix~\ref{sec:cataloguemanual}\footnote{By \textit{Kron-like isophote} we refer to the Kron ellipse enlarged by a factor of 1.5.}. This information is now easily accessible during the parametric fitting routine and helps to make decisions on the models to be used to simultaneously fit the objects lying in each stamp (see section~\ref{sec:galfit}); indeed, they are crucial also to the non-parametric approach, since they communicate to ZEST+ all the necessary information to clean the neighbours in the stamps and prepare them for the measuring routine which is described in section~\ref{sec:zestp}.

\section{Parametric Fits}
\label{sec:tools}
\subsection{\textsc{Galfit} Setup}
\label{sec:galfit}
Image cutouts and PSF models appropriate to each individual object are provided to \textsc{Galfit}, which is used to find the best-fitting S\'ersic models. As reported in  \citep{Peng124,Peng2010}, the adopted S\'ersic function has the following form: 
\begin{equation}
\Sigma(r) = \Sigma_e \exp \bigg\{ -k \bigg[ \bigg(\frac{r}{R_e}\bigg)^{\frac{1}{n}} -1 \bigg] \bigg\},
\label{eq:sersicfunction}
\end{equation}
where $\Sigma_e$ is the pixel flux at the half-light radius $R_e$. The S\'ersic index \textit{n} quantifies the profile concentration: if \textit{n} is large, we have a steep inner profile with a highly extended outer wing; inversely, when \textit{n} is small, the inner profile is shallow and presents a steep truncation at large radii. In the case of $n=1$ we have an exponential light profile. We indicate with $k$ the normalization constant coupled to the S\'ersic index so that the estimated effective radius always encloses half of the flux (elsewhere, $b_n$ is sometimes used for this quantity). \textsc{Galfit} produces measurements for the free parameters of the S\'ersic function: central position, integrated magnitude ($m_{tot}$), effective radius ($R_e$) measured along the major axis, S\'ersic index (\textit{n}), axis ratio (\textit{q}) and position angle (\textit{PA}). The integrated magnitude is determined through its definition as a function of the flux ($F_{tot}$) integrated out to $r=\infty$ for the S\'ersic profile:
\begin{equation}
m_{tot} = -2.5\log\bigg(\frac{F_{tot}}{t_{exp}}\bigg)+mag\_zpt,
\label{eq:integratedmagnitude}
\end{equation}
where $t_{exp}$ is the exposure time and $mag\_zpt$ is the zero-point magnitude, both indicated in the image header.

Apart from the central position, which is allowed to vary by only $\pm 1$ pixel by a \textsc{Galfit} constraints file, all the parameters are left free without constraints: for those, initial guesses are taken from the \textsc{SExtractor} DESDM catalogues (the exception being S\'ersic index, which is always started at $n=2$ and, according to our tests, produces negligible fluctuations in the output if started at other values). Thanks to the large background area available in each stamp (pre-validated with the SAND algorithm), \textsc{Galfit} is left free to estimate the background level\footnote{During initial tests on the fitting routine we randomly selected a sub-sample of objects to be fitted with the background fixed to zero. The outcome of this test was that this choice does not change significantly the results. 
}. \\
For the measurements, \textsc{Galfit} is left free to build the sigma-image internally.
We explored different sizes of the cutout images and convolutions boxes, sequentially enlarging the image until convergence was achieved. Given $X$ and $Y$ the dimensions of the cutout image (in pixels), we set the dimensions of the convolution box to $(X+2, Y+2)$	pixels. \\
The information provided by the SAND routine is adopted in order to optimise the simultaneous fitting procedure of the central galaxy and its neighbours. Using the \texttt{ELLIPSE\_FLAGS} (introduced in section~\ref{sec:prefittingpip}) it is easy to identify most of the neighbours, including faint companions, nearby stars, close objects with overlapping isophotes and neighbours with unreliable priors due to unphysical \textsc{SExtractor} measurements. \\
Companion objects three magnitudes fainter than the main galaxy are not fit. In the presence of overlapping isophotes, the relevant neighbouring object is fit simultaneously with the target galaxy (even in the cases where it is centred outside the stamp). However, if the overlapping region is $50 \%$ or larger than the area within the Kron-like ellipse occupied by the central galaxy, then although a fit is attempted, it is not considered for the analysis discussed in this paper.
Given $k1$ and $k2$ as the effective Kron Radii of the central galaxy and its neighbour respectively, they are used to define the isophotes of those objects, intended as enlarged Kron-like ellipses. If the isophotes are not overlapping, but separated by less than the maximum between $k1$ and $k2$, then the neighbour is fit simultaneously. Otherwise it is masked. If the neighbour is a star ($\texttt{CLASS\_STAR}\ge0.9$), it is simultaneously fit with a PSF model. Finally, if the stamp contains one or more neighbours whose initial guesses from \textsc{SExtractor} contain errors (for example negative magnitudes and radii), no fit is attempted. We adopt a Single S\'ersic model with all its parameters free for neighbours also.

\begin{figure*}
	\includegraphics[width=\textwidth]{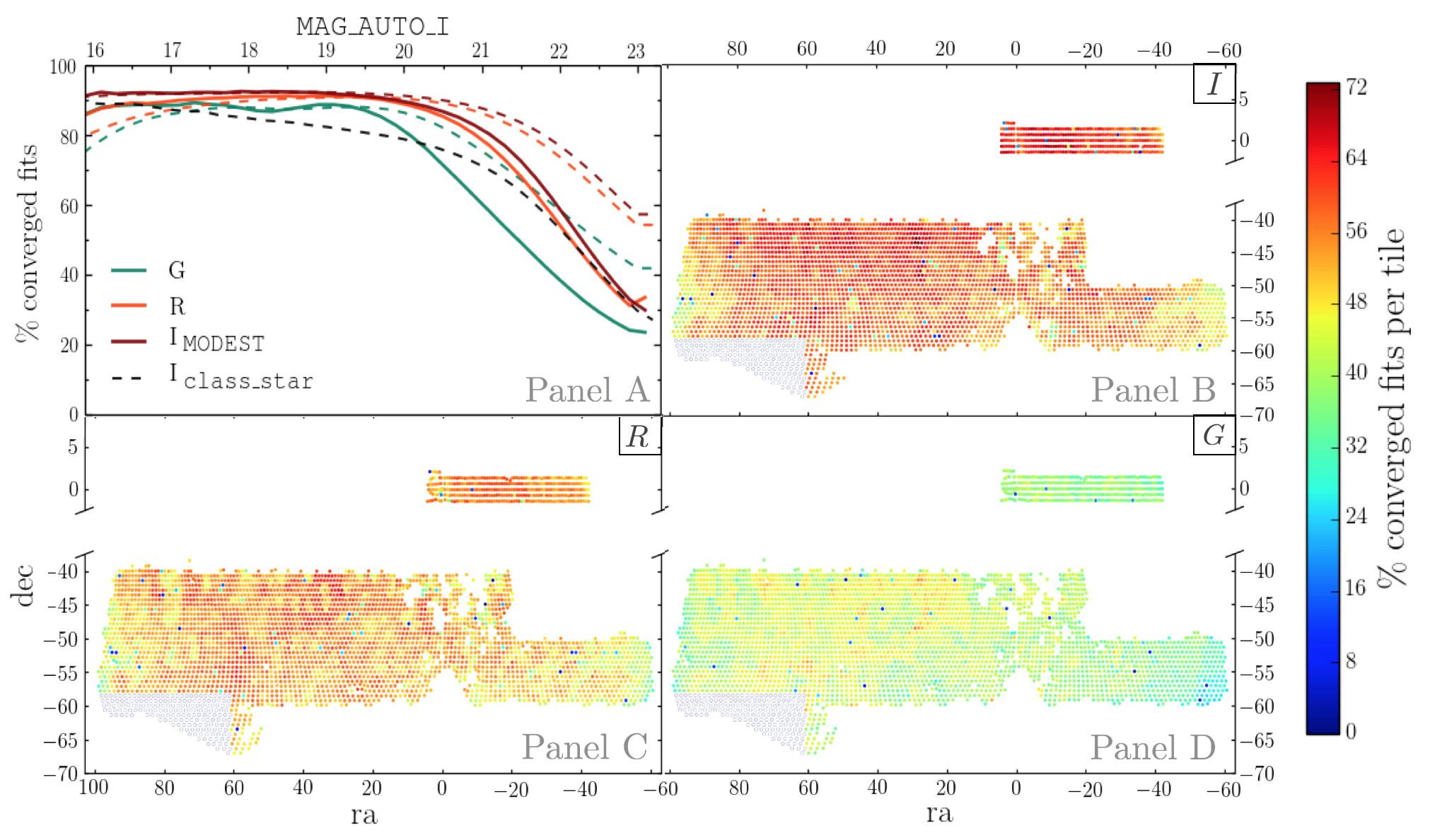}
	\caption{\textbf{Panel A}: fitting completeness in \textit{g},  \textit{r} and \textit{i} bands (green, orange and brown lines, respectively), following star-galaxy separation using the \texttt{MODEST} classifier (see Section~\ref{sec:stellarcontamination}). The completeness, defined in eq.~\ref{eq:compl_eq}, is expressed in terms of the percentage of converged fits calculated in bins of 0.2 magnitude. Solid lines show the completeness in differential magnitude bins, while the dashed lines show results for magnitude-limited samples. The dashed black line shows the trend for the \textit{i} band when using only a conservative S/G cut ($\texttt{CLASS\_STAR} > 0.9$). Using the \texttt{MODEST} classifier we find that the completeness is $~90 \%$ up to magnitude 21.
		\textbf{Panels B, C, D: } maps of the percentage of converged fits in \textit{g}, \textit{r} and \textit{i} band in each tile (at $mag\_auto\_i < 23$). The region in the lower left corner occupied by empty grey circles is entirely flagged as unsuited for extra-galactic work due to its vicinity to the Large Magellanic Cloud (LMC). The regions with a lower fraction of converged fits are found towards the Galactic Plane and close to the LMC. In the \textit{g} band the percentage of converged fits is poorer, as expected, due to an overall broader PSF.
	}
	\label{fig:convergedmap}
\end{figure*}

\begin{figure*}
	\includegraphics[width=\textwidth]{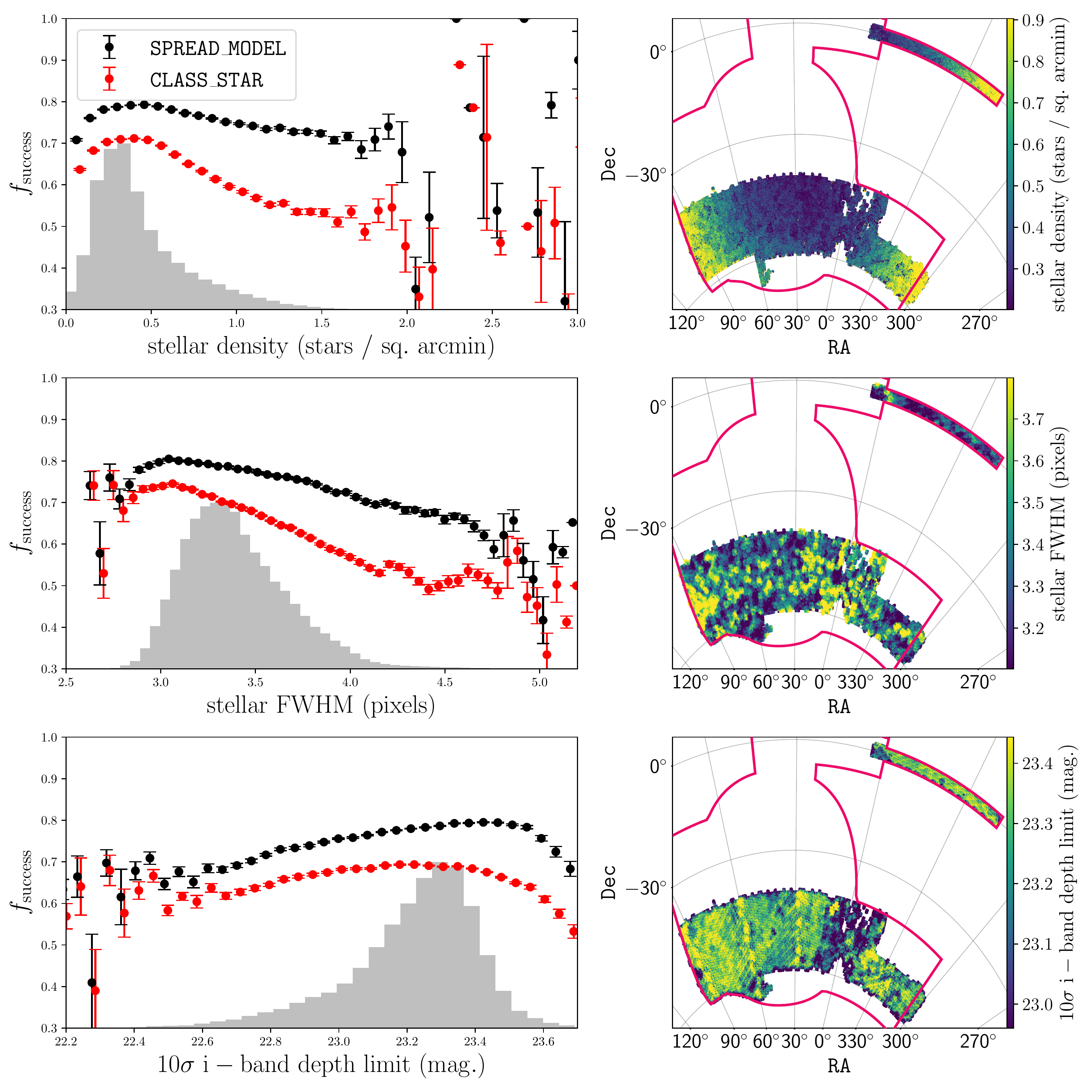}
	\caption{Dependence of fitting completeness at $i<21.5$ on spatially-dependent survey characteristics, stellar density, PSF FWHM and \textit{i}-band image depth (top, middle and bottom panels respectively). The maps of the nominal DES five-years footprint (outlined in magenta) show the dependences for the DES Y1 area. Grey histograms show the relative distributions of the characteristics in terms of survey area. The results for the galaxy sample are shown, following two star-galaxy classifiers: \textsc{SExtractor} \texttt{CLASS\_STAR} (red points) and an additional criterion based on \texttt{SPREAD\_MODEL} (black points, see text). Uncertainties are derived by bootstrap resampling. After the improved S/G separation, the fitting completeness is only weakly dependent on survey characteristics, and a high completeness ($>80\%$) can be maintained with only minimal loss of area. The results at $i<22$ are very similar in terms of the correlations with survey characteristics, but with overall lower converged fraction. 
	}
	\label{fig:systematics}
\end{figure*}

\begin{figure*}
	\includegraphics[width=\textwidth]{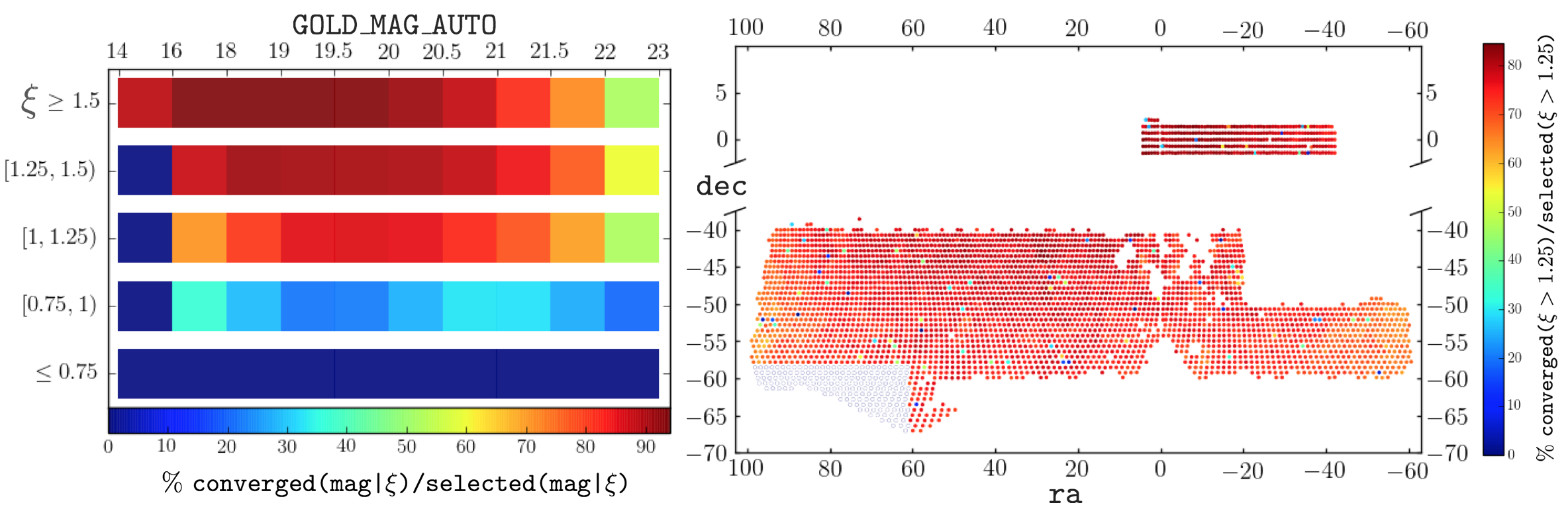}
	\caption{\textbf{Left panel}: fitting completeness calculated in differential bins of magnitude. The sample is divided into sub-populations, according to different ranges of the parameter $\xi = \texttt{FLUX\_RADIUS/PSF\_radius}$, as reported on the y-axis. Each population is represented by a bar, colour-coded by the percentage of converged fits in each magnitude bin. The figure shows that failed fits are more frequent for the objects with size smaller than the PSF or comparable with it. A critical drop occurs for the population with $\xi<1.25$. \textbf{Right panel: }map of the percentage of converged fits per tile with $\xi>1.25$. In comparison with the \textit{i} band map in Fig.~\ref{fig:convergedmap}, it is clear that by applying this cut the overall percentage of successful fits increases dramatically, from $\sim40 \%$ to $> 70 \%$ at the borders and up to $\sim 90 \%$ in the central areas.}
	\label{fig:ratio}
\end{figure*}

\begin{figure*}
	\includegraphics[width=\textwidth]{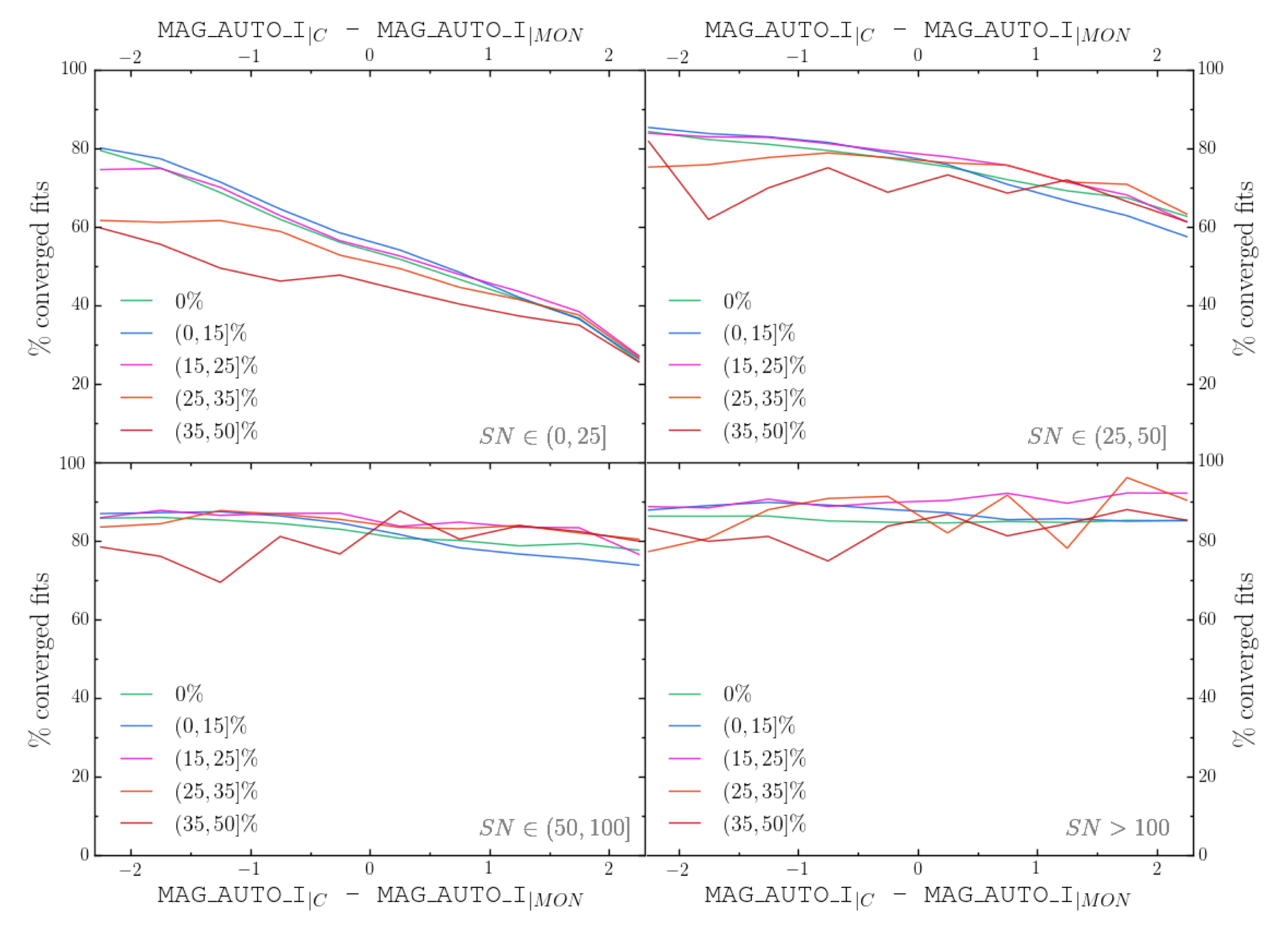}
	\caption{Fitting completeness as a function of the magnitude difference between the target galaxy and its closest neighbour. The relation is shown for different percentages of overlap between the two fitted objects, as reported in the legend. Each line is normalized by the population of objects with attempted fits within the same range in magnitude difference. The analysis is repeated in four signal-to-noise intervals. We observe that the fitting completeness decreases when the most overlapping neighbour is much brighter than the central galaxy, with stronger effects in low signal-to-noise regimes. This effect becomes negligible with increasing signal-to-noise.}
	\label{fig:compl_neighbours}
\end{figure*}

\subsection{Fitting Completeness}
\label{sec:parametric_completeness}
\textsc{Galfit} uses a non-linear least-squares algorithm which iterates $\chi^2$ minimization in order to find the best solution given a large parameter space.  However even when the algorithm outputs a solution, there could be cases where the estimation of one or more parameters is affected by numerical convergence issues, which makes the solution itself an unreliable and non-unique result. These cases include correlated parameters, local minima and mathematically degenerate solutions (\citealt{Peng2010}, Section 6). \textsc{Galfit} labels the affected parameters enclosing them in between stars ($**$). In such cases we classify the fit as non-converged and do not trust the set of structural parameters it provides. \\ 
We determine the fraction of converged and non-converged fits and investigate their properties and location in the DES field. We present our analysis for all filters taking the \textit{i} band as reference to discuss the fitting properties and possible causes of failure and incompleteness.\\
\color{white} text \color{black} \\
We evaluate the fitting completeness by calculating the percentage of converged fits in differential bins of 0.2 magnitude. The completeness ($\mathcal{C}$) is calculated by normalising the number of converged fits in each magnitude bin ($N(c|mag)$) to the number of objects which passed the sample selection (described in section~\ref{subsec:sampleselection}) in that bin, as expressed in the following definition:
\begin{equation}
\label{eq:compl_eq}
\mathcal{C}_{|mag} = \frac{N(c|mag)}{N(c|mag)+N(nc|mag)+N(f|mag)},
\end{equation}
where $N(nc|mag)$ and  $N(f|mag)$ refer to the fractions of non-converged and failed fits in each magnitude bin, respectively. We also derive the percentage of converged fits calculated within limiting magnitudes. \\
\color{white} text \color{black} \\
The results of this analysis are shown in Fig.~\ref{fig:convergedmap}. In the upper left inset (Panel A) the solid lines represent the fitting completeness in magnitude bins and the dashed lines the magnitude limited completeness. They are colour-coded by filter: green and orange lines refer to \textit{g} and \textit{r} band, respectively; brown and black to the \textit{i} band. We start our discussion from the latter.\\
The dashed black line shows the completeness determined for a sample with a conservative star-galaxy (S/G) cut ($\texttt{CLASS\_STAR} > 0.9$): the trend shows that $\sim90\%$ of the fits are successful at magnitude $\sim17$, after which this value starts to decline and reaches $\sim80\%$ at magnitude $\sim21$. The completeness decreases more rapidly towards fainter magnitudes. \\ 
The brown line shows the completeness after applying a star-galaxy cut based on the \texttt{SPREAD\_MODEL} parameter (further details on the star-galaxy classifier and associated analysis are described in the following subsection). In this galaxy sample, a completeness of  $\sim85\%$ is reached at magnitudes $I<21.5$.
We match the information given by the first panel with the map in Panel B: each point represents a DES tile and is colour-coded by the percentage of converged fits in that tile. The area identified by empty grey circles, where $100<ra<60$ and $-70<dec<-58$, has been excluded from the sample selection because in the \texttt{GOLD} catalogue it is flagged due to its vicinity to the Large Magellanic Cloud (LMC).\\
We observe that the regions with a higher percentage of non-converged fits are located at the East and West borders of the footprint, towards the Galactic plane. These regions are characterized by high stellar density, as shown in \cite{Pieres}. One possibility is that many of the unconverged fits at relatively bright magnitudes are stellar contaminants and so there is a poorer completeness where the stellar spatial frequency is higher. Another scenario could be that the edges of the footprints were observed under poorer conditions, for instance with poorer seeing. We now investigate these possibilities thorugh examining correlations between our fitting completeness and maps of survey characteristics (as introduced in \citealt{Leistedt} and \citealt{Drlica-Wagner}), and discuss the likely causes of failures, encompassing stellar contamination, the effect of PSF width, poor signal-to-noise and the effects of neighbouring sources.

\subsubsection{Stellar contamination}
\label{sec:stellarcontamination}
We used the neural network star-galaxy (S/G) classifier, included as part of \textsc {SExtractor}, for a conservative initial criterion of star-galaxy separation. We apply the cut $\texttt{CLASS\_STAR} < 0.9$, in order to remove only the most obvious stars, and to allow a user to perform their own S/G separation.
Point sources will most likely fail to achieve a converged solution in \textsc{Galfit} and we therefore expect that a substantial fraction of the incompleteness at bright magnitudes seen in the black dotted line in Fig.~\ref{fig:convergedmap} (panel A) is due to contamination by stellar sources. This expectation is supported by the fact that the regions with the lowest percentage of converged fits (Fig.~\ref{fig:convergedmap}, panels B-D) are located in regions of known high stellar density. Further, in the upper panel of Fig.~\ref{fig:systematics} it can be seen that the converged fraction at $i<21.5$ depends strongly on the stellar density for the \texttt{CLASS\_STAR} S/G separation.

In \cite{Drlica-Wagner} it is shown that a simple cut in the \textsc{SExtractor} parameters \texttt{SPREAD\_MODEL} and \texttt{SPREADERR\_MODEL} can achieve a galaxy completeness of $\ge98\%$, with $\le3\%$ stellar contamination at $i<22$. This cut is known as \texttt{MODEST} classifier.
\texttt{SPREAD\_MODEL} is a morphological quantity which compares the source to both the local PSF and a PSF-convolved exponential model \citep{Desai, Soumagnac2015}. In order to optimise the separation of point-like and spatially extended sources, we use the \textit{i} band as the reference band for object classification due to the depth and superior PSF in this filter. The separation is defined via a linear combination of the \texttt{SPREAD\_MODEL} and its uncertainty, the \texttt{SPREADERR\_MODEL}:
\begin{equation}
\label{eq:sgalsep}
\texttt{SPREAD\_MODEL} + n \times \texttt{SPREADERR\_MODEL} > thr,
\end{equation}
where the coefficients $n=1.67$ and $trh=0.005$ are chosen as the optimal compromise between the completeness and purity of the galaxy sample. With the \texttt{MODEST} classifier we recover more than $\sim90\%$ converged fits at magnitude 20 and $\sim85\%$ at magnitude 21.5.\\
We apply this additional S/G classification henceforth, and show the converged fraction of galaxies under this additional classification by the coloured lines in Fig.~\ref{fig:convergedmap} and the black points in Fig.~\ref{fig:systematics}. The dependence of converged fraction on stellar density is vastly reduced with the \texttt{SPREAD\_MODEL} classifier (though still present) with a threefold increase in stellar density, from $0.5$ to $1.5$ stars per sq. arcmin, causing just a $7\%$ point drop in converged fraction. This decrease is almost entirely explained by the expected contamination rate of $3\%$.

\subsubsection{PSF width}
\label{sec:PSFeffect}
In order to take into account the seeing, \textsc{Galfit} convolves the 2-D model with the PSF, and compares it with the galaxy image. For this reason galaxy fitting requires very accurate knowledge of the PSF. Errors in the PSF model can easily result in attempted fits not converging, or in biased parameters (see section \ref{subsec:paramsimulations}). Here, we assess the fitting incompleteness due to the varying PSF width across the DES survey area. We calculate the completeness for different sub-populations of the sample, delimited by certain values of the ratio between the galaxy half-light radius, estimated by the \textsc{Sextractor} \texttt{FLUX\_RADIUS}, and the PSF size; we indicate this parameter with $\xi$, defined as follows:
\begin{equation}
\xi = \frac{\texttt{FLUX\_RADIUS} }{\texttt{PSF\_radius}},
\label{eq:R}
\end{equation}
where we calculate the size of the PSF as the radius of the circular aperture enclosing half of the flux of the PSF itself. The typical PSF radius is $\sim 3 \ px$. The left panel in Fig.~\ref{fig:ratio} shows the completeness calculated in bins of 1 magnitude for five different populations: $\xi\le0.75$, $0.75<\xi\le1$, $1<\xi\le1.25$, $1.25<\xi\le1.5$ and $\xi\ge1.5$. Values of $\xi<1$ are unphysical, indicating either noisy photometry, image artefacts or inaccuracies in the PSF model. Each population is represented with a bar coloured by the percentage of converged fits, normalised by the total number of selected objects in each magnitude bin. As expected, we observe lower percentages of converged fits for the objects whose size is comparable to the size of the PSF used by \textsc{Galfit} to deconvolve their images. Nevertheless, in the range $1<\xi\le1.25$ the completeness is only around $10\%$ lower than at larger sizes. The right panel in Fig.~\ref{fig:ratio} maps the completeness per tile, excluding the galaxy sample whose size is comparable or smaller than the PSF ($\xi<1.25$). Compared with the \textit{i} band map in Fig.~\ref{fig:convergedmap}, it shows that by applying the cut in $\xi$ the fitting completeness increases dramatically both at the borders (up to $>70 \%$) and in the central areas (up to $\sim 90 \%$), and the discrepancy between these two regions is reduced. \\
In Fig.~\ref{fig:systematics}, centre panel, we show the dependence of fitting completeness against PSF FWHM ($i<21.5$). For the \texttt{SPREAD\_MODEL} S/G classifier we see that the completeness at $i<21.5$ only drops below $80\%$ in the extended tail of the distribution of PSF FWHM (grey histogram). 

\subsubsection{Image depth}
\label{sec:deptheffect}
There is a clear and expected dependence of the percentage of converged fits on magnitude in both Fig.~\ref{fig:convergedmap} and Fig.~\ref{fig:ratio}. Although stars are less easily excluded at faint magnitudes and the sizes of galaxies are smaller, much of this dependence is likely to be due simply to the difficulty of \textsc{Galfit} finding a stable minimum in the $\chi^2$ space at low S/N. In the lower panel of Fig.~\ref{fig:systematics} we show how the fitting success rate for $i<21.5$ galaxies depends on image depth, and hence object S/N. As expected, the completeness falls in shallower regions of the footprint, but the decline is not dramatic for this bright subset and, once again, a high success rate can be maintained by removing only regions corresponding to the tails of the distribution.

\subsubsection{Impact of neighbouring sources}
\label{sec:neighbour}
Finally, we assess the impact of neighbouring sources on the fitting success rate. We reduce the complexity of possible arrangements of neighbours to two metric values: the amount of overlapping area\footnote{By area, we mean the \textsc{SExtractor}-derived Kron ellipse enlarged by a factor of 1.5} between a galaxy and its neighbours, and the difference in magnitude between the galaxy and its most overlapping neighbour ($(\texttt{MAG\_AUTO\_{|C}})-(\texttt{MAG\_AUTO\_{|MON}})$). The dependence of the converged percentage as a function of these two quantities is shown in Fig.~\ref{fig:compl_neighbours}, in four intervals of S/N for the target object. Each line in the figure is normalized by the population of objects with attempted fits within the same delta-magnitude range.
We observe that even at low S/N the fitting success rate is high if all the neighbours present are sufficiently faint. However, in the range $0<S/N<25$ the completeness is a steep function of the magnitude difference between target galaxy and its neighbour. At high S/N neither the degree of overlap nor the relative magnitude of a neighbour are important. Note that, our initial selection removes objects that \textsc{SExtractor} determined to have been blended.

\subsubsection{Multi-wavelength completeness}
As shown by the green and red curves in Panel A in Fig.~\ref{fig:convergedmap}, we can recover a relatively high percentage of converged fits for objects brighter than magnitude 21.5 for the \textit{g} and \textit{r} filters also. We notice that the \textit{g} and \textit{r} bands show a drop in the brightest magnitude range ($\texttt{GOLD\_MAG\_AUTO\_i} \le 15.5$). Upon inspection we find that the objects responsible are compact objects with size comparable to the PSF and with a \texttt{MODEST} classification which is close to the threshold of 0.005 in the \textit{i}-band. 
In Panels C and D we can see the spatial completeness for the \textit{r} and \textit{g} band, respectively. In both cases we reconfirm what we observed for the \textit{i} band: a poor fitting completeness at the borders of the field, where stellar density is high, as discussed in the previous sub-sections. The \textit{g} band PSF is typically broader then the \textit{r} and the \textit{i} bands, and the images shallower, which are reflected in an overall poorer recovery of converged fits.

\subsection{Validation}
\label{sec:validation}
\begin{figure}
	\includegraphics[width=\columnwidth]{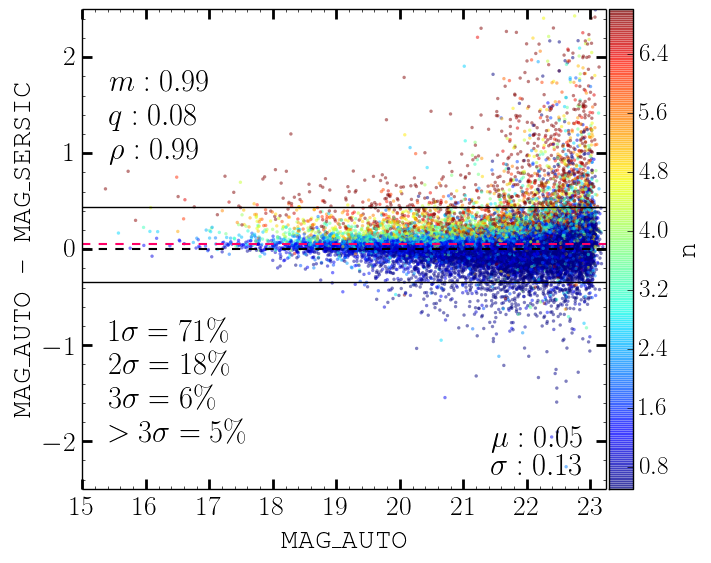}
	\caption{Difference between the input magnitude (\texttt{MAG\_AUTO}) from \textsc{SExtractor} and the output magnitude (\texttt{MAG\_SERSIC}) recovered through Single S\'ersic fits. Results are shown as a function of input magnitude and are colour-coded by S\'ersic Index. The two solid black lines delimit the population lying within 3 standard deviations from the mean magnitude difference relation, indicated by the dashed red line. The mean and the spread of the relation, printed in the lower right corner of the Figure, are obtained through a $3\sigma$ clipping procedure. The banding in S\'ersic index is expected \citep{Graham05} and the vast majority of outliers (which in total number $5\%$ of the sample) are of low S/N objects.}
	\label{fig:magmag}
\end{figure}

We now turn to assessing the accuracy of the parameters recovered from those objects that were successfully fit with \textsc{Galfit}, beginning with simple magnitude and size diagnostics of the population. We then investigate whether there are systematic errors from which \textsc{Galfit} suffers in recovering the structural parameters of the galaxies, depending on their magnitude, size, concentration and shape. We investigate this aspect through image simulations (section~\ref{sec:simulations}) and present the relative calibrations in the next subsection.

For this discussion we show the tests performed on the \textit{i} band, which represents our fiducial filter, starting with a comparison of the total S\'ersic magnitude with \texttt{MAG\_AUTO} computed by \textsc{SExtractor}. In Fig.~\ref{fig:magmag} we show this comparison for 30,000 randomly-selected objects from the full catalogue. We recover the expected behaviour: objects with S\'ersic index $\sim1$ have magnitudes consistent with \texttt{MAG\_AUTO}, while the S\'ersic magnitude is brighter at higher $n$. \texttt{MAG\_AUTO} is known to be biased faint for high-S\'ersic $n$ objects, losing as much as $50\%$ of the flux in extreme cases \citep{Graham05}.

The solid black lines in Fig.~\ref{fig:magmag} delimit the $3\sigma$ outliers in magnitude difference, following an iterative 3-sigma-clipping procedure to find the mean relation and spread (given by the parameters, $\mu$ and $\sigma$ in the figure). The mean relation (red dashed line) is essentially flat in magnitude, suggesting that typically the background computed during catalogue extraction and that estimated by \textsc{Galfit} are consistent. At faint magnitudes, however, there is a population of outliers with magnitude differences that cannot be explained by simple photometric errors, and that also exhibit very high S\'ersic indices. We deem these unreliable fits, possibly caused by an unidentified elevated background. Restricting the sample to objects with $S/N>30$ removes these objects and entirely removes the group with spurious large radii.

\begin{figure}
	\includegraphics[width=\columnwidth]{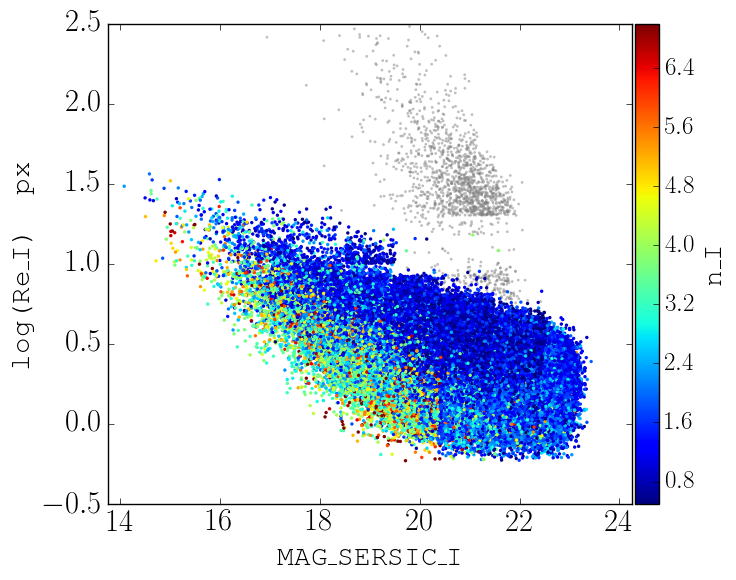}
	\caption{Relation between S\'ersic magnitude and effective radius for the \textit{i} band results. Points are colour-coded by S\'ersic Index. outliers are shown in grey.}
	\label{fig:magsize_plane_uncalib}
\end{figure}

We then obtain the relation between magnitude and effective radius from the S\'ersic profile fits as shown in Fig.~\ref{fig:magsize_plane_uncalib}. Points are colour coded by each object's S\'ersic index. Once again, the data match expectations and similar trends reported in the literature, with high S\'ersic $n$ objects forming a steep sequence and galaxies with exponential light profiles dominating at fainter magnitudes. Grey points are sources labelled as outliers during the validation process.

\begin{figure}
	\includegraphics[width=\columnwidth]{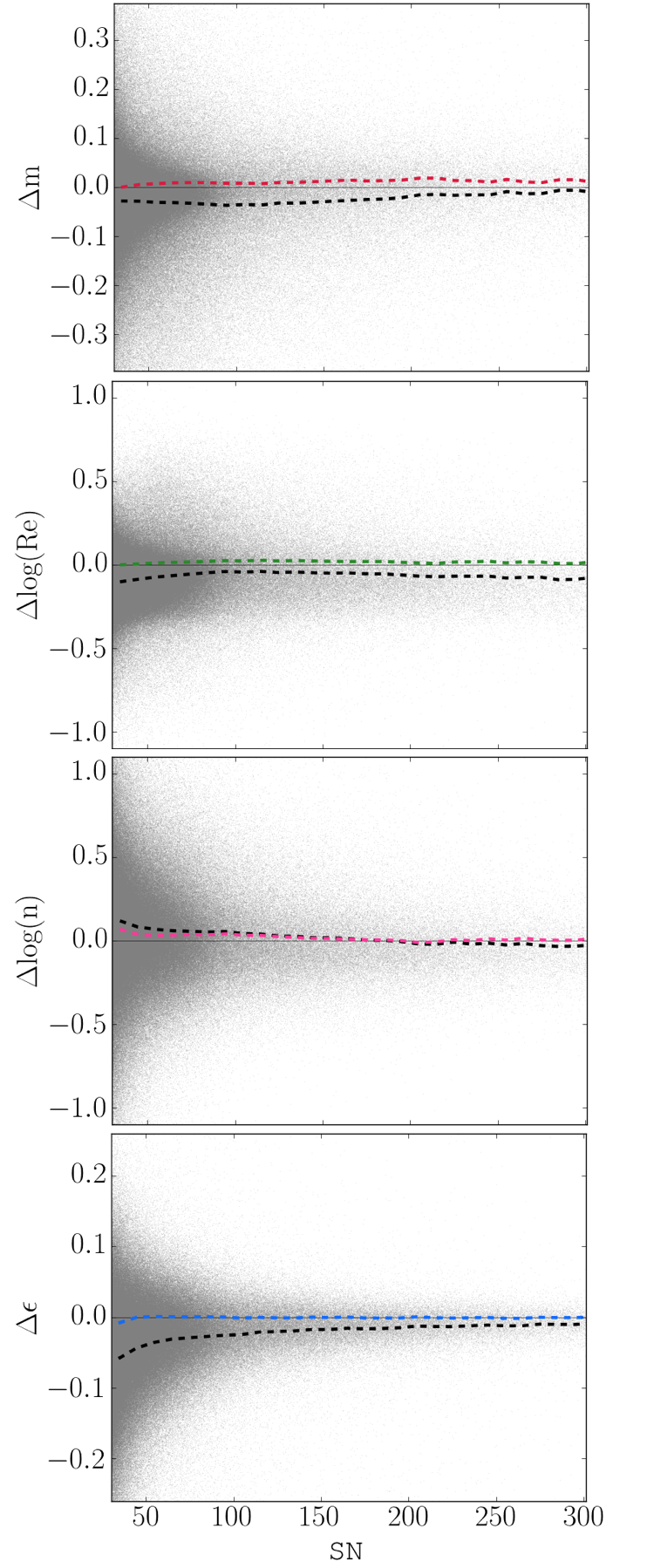}
	\caption{Discrepancies in recovered S\'ersic parameters from running \textsc{Galfit} on the \textsc{UFig-BCC} image simulations, as a function of signal to noise (S/N). From top to bottom the panels display the results for magnitude, half light radius, S\'ersic index and ellipticity. The dashed lines show the discrepancy in bins of S/N, calculated before (black line) and after (coloured line) applying calibration corrections (see section \ref{subsec:paramsimulations}). The uncertainties depend to first order on the signal to noise, and the mean deviation is clearly reduced by applying the calibrations. In the calibration map, shown in figure \ref{fig:mapcalsnew}, we investigate how the parameters and their uncertainties correlate with each other.}
	\label{fig:cals_sn}
\end{figure}

\begin{figure*}
	\includegraphics[width=\textwidth]{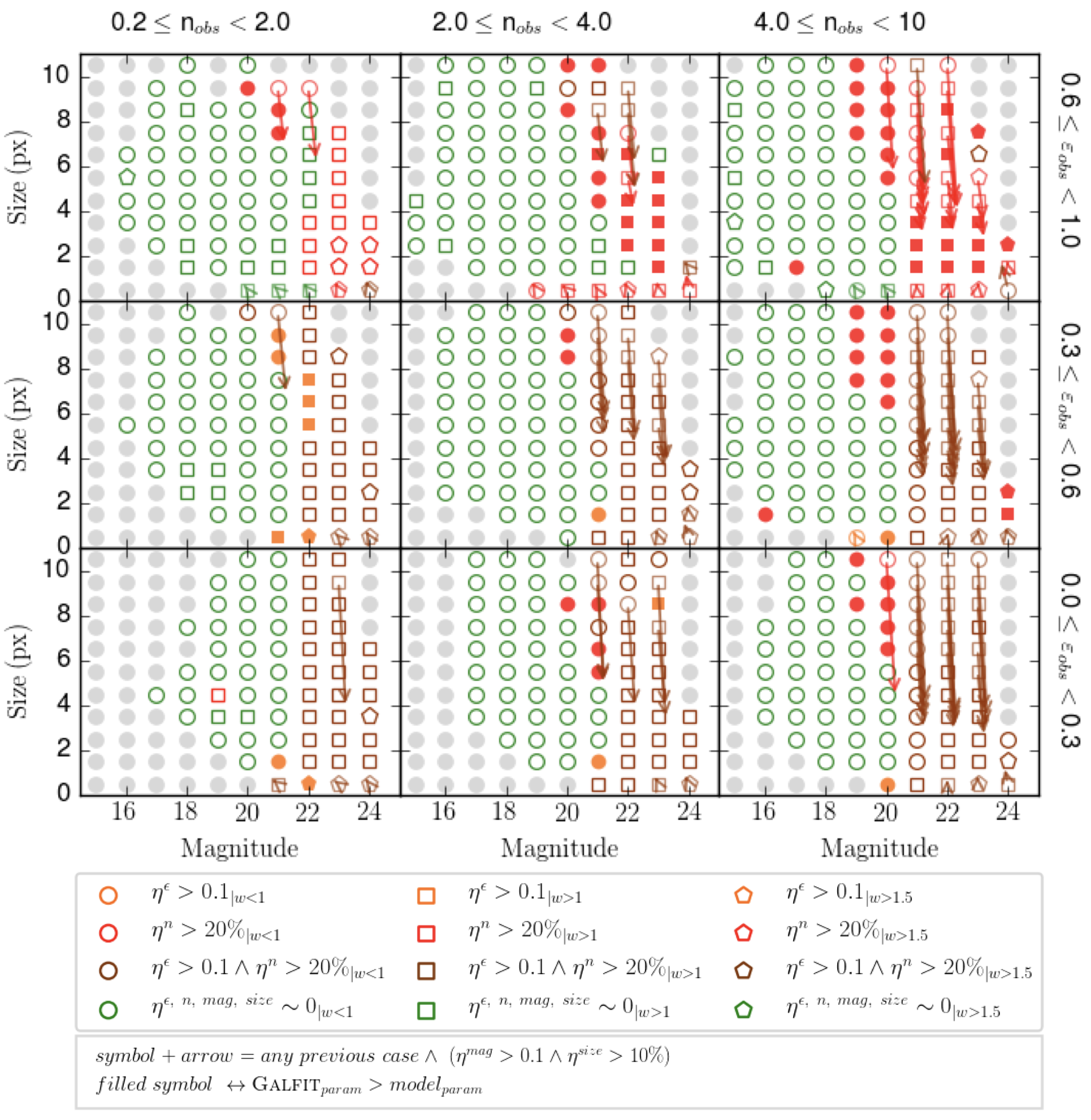}
	\caption{Calibration map for the parametric measurements in the \textit{i} band, obtained from image simulations as described in Section~\ref{sec:simulations}. The calibrations are determined in a 4D parameter space, where the correlation of size, magnitude, ellipticity and S\'ersic Index between the simulated galaxy and the model is studied. The information is provided using different marker shapes (circles, squares, pentagons, arrows) and colours, as follows.  The calibrations are presented in a size-magnitude plane, divided in different cells according to the shown sub-ranges in ellipticity and S\'ersic Index. The components of the correction vectors are the magnitude discrepancy $\eta^{mag}$ and the size discrepancy $\eta^{size}$, according to the definitions given in Equations~\ref{eq:calsmodel} and~\ref{eq:calsfits}. If these corrections are small ($\eta^{mag} < 0.1  \wedge \  \eta^{size}<10 \%$) the length of the arrow is set to zero and the cell is identified by a symbol only. Points and arrows are coloured according to the scatter in ellipticity ($\epsilon$) and S\'ersic Index ($n$); a scatter in $\eta^{\epsilon}>0.1$ or $\eta^{n} > 20\%$ is expressed in orange and red, respectively, while the cells presenting a large scatter in both parameters are coloured in brown. The symbol is empty if the \textsc{Galfit} recovered value is smaller than the model. Different shapes are used referring to the total scatter (w) in the 4D parameter space of the model parameters, defined in Equation~\ref{eq:variance}; the symbol is a pentagon if $w>1.5$ and a square if $w>1$, otherwise it is a circle. The symbols and conventions used in the calibration map are summarised in the legend. In the case of the calibration of non-parametric fits (following in Section~\ref{subsec:npsimulations}) the S\'ersic Index is replaced with the concentration parameter.}
	\label{fig:mapcalsnew}
\end{figure*}

\subsection{Calibrations}
\label{subsec:paramsimulations}
In this section we illustrate how we calibrate our measurements. As explained in detail in Section~\ref{sec:simulations}, we processed and fit the \textsc{UFig-BCC} simulated data for DES Y1 in the same way we did for our real galaxy sample. We used $\sim 10$ million simulated objects. Now we can compare the results from the fits with the true morphological parameters used to generate the \textsc{UFig-BCC} images. We then calculate the discrepancies between the measured and true parameters and derive appropriate corrections. We show the size of these corrections via a set of calibration maps.

\subsubsection{Derivation of the corrections}
\label{subsec:correction_calc}
We derive corrections in a 4-dimensional parameter space, including size, magnitude, S\'ersic Index and ellipticity. The ensemble of values assumed by each parameter constitutes a vector in the parameter space. We sample each vector with a list of nodes: the magnitude ($mag$) in the range [14.5,23.5] in steps of 1 magnitude, the size ($r$) in the interval [0.5,16.5] px in steps of 2 px, the S\'ersic Index ($n$) in the set [0.2, 2, 4, 10] and the ellipticity ($\epsilon$) in the intervals [0, 0.3, 0.6, 1]. The realization of each combination of these nodes forms an hypervolume which we'll refer to as a \textit{cell}. In each cell falls a certain number of simulated objects with similar structural properties and the corresponding fitting results: so each parameter is represented by a distribution of simulated values and a distribution of measurements. Each distribution in turn has a median value ($m^{i}$) and a standard deviation ($\sigma^{i}$), where $i={mag, r, n, \epsilon}$, which represent the central value and  the dispersion of the population, respectively. To summarise, in each cell the i-th parameter can be expressed as:
\begin{equation}
\hat{i} = \hat{\mu}^{i} \pm \hat{\sigma}^{i}
\label{eq:calsmodel}
\end{equation}
for the model and as:
\begin{equation}
i = \mu^{i} \pm \sigma^{i},
\label{eq:calsfits}
\end{equation}
for the fit, where $i (\hat{i})={mag, r, n, \epsilon}$. For all the objects falling in a given cell we calculate the correction ($\eta^{i}$) in each parameter as the discrepancy between the central values of the distributions:
\begin{equation}
\eta^{i} =  \hat{\mu}^{i} - \mu^{i}. 
\label{eq:cals}
\end{equation}
We further define a quantity, $w$, which represents the dispersion of the cell in the 4D parameter space, derived as the quadratic sum of the variances of the model parameters which determine the diagonal of the covariance matrix of the parameter space. It is defined as follows:
\begin{equation}
w = \sqrt{ \sum_{i} \frac{ \hat{\sigma_{i}}^{2}}{\hat{m_{i}}^{2}} },
\label{eq:variance}
\end{equation}
where $i ={mag, r, n, \epsilon}$ and $\hat{\sigma_{i}}^{2}$ and $\hat{m_{i}}$ are the variance and median values of the model distributions, respectively. For cells with larger dispersion, we expect the correction vector to be less accurate for a given randomly chosen object. 

\subsubsection{Calibration maps}
In the validation routine we observed that $\sim 99 \%$ of converged fits are well recovered in magnitude ($\eta^{mag}$ of the order of 0.001), and that cutting objects with $S/N<30$ we remove the clear outliers in size and magnitude. In Figure~\ref{fig:cals_sn} we show the discrepancies $\eta^{i}$ between the intrinsic values and the parametric measurements as a function of signal to noise for magnitude, half-light radius, ellipticity and S\'ersic index. The discrepancies relative to size and S\'ersic index are shown in logarithmic space to facilitate visualization. In each panel the dashed lines show the discrepancies in bins of signal to noise. We use the uncalibrated sample to calculate the black line, and the same sample after applying the calibrations for the coloured one. It is clear that the uncertainties on the structural parameters increase in low signal to noise regimes, as one might anticipate, and the scatter clearly reduces when applying the corrections. We observe that \textsc{Galfit} tends to recover larger sizes and ellipticities, so we pay particular attention to the corrections required for these properties within the multidimensional parameter space.

Figure~\ref{fig:mapcalsnew} represents a map of the calibrations that we apply to our measurements, derived from our state-of-the-art image simulations. In using this multidimensional calibration map we are able to account for the correlations between parameters and ensure the corrections are appropriate for a true galaxy sample.
The arrows represent the strength of the vector corrections, expressed as the distance between the central values of the size and magnitude distributions of the model sample and the relative measured dataset in each cell. The components of the correction vectors are the magnitude discrepancy $\eta^{mag}$ on the x axis and the size discrepancy $\eta^{size}$ on the y axis, according to the definitions given in Equations~\ref{eq:calsmodel} and~\ref{eq:calsfits}. If these corrections are small ($\eta^{mag} < 0.1  \wedge \  \eta^{size}<10 \%$) the length of the arrow is set to zero and only a circle is shown. Apart from the grey circles, which indicate areas with poor statistics, different colours are used to give an indication of the correction applied to ellipticity and S\'ersic Index. If $\eta^{\epsilon}>0.1$ or $\eta^{n} > 20\%$, the symbol is coloured in orange and red, respectively. If the correction is large in both cases, then it is coloured in brown. The symbol is empty if the \textsc{Galfit} recovered value is smaller than the model. The symbols are shaped according to the total scatter (w) in the 4D parameter space of the model parameters, defined in Equation~\ref{eq:variance}; we use a pentagon if $w>1.5$ and a square if $w>1$, otherwise the symbol is a circle. Figure~\ref{fig:mapcalsnew} reports the vector corrections for the \textit{i} band; corrections for the \textit{g} and \textit{r} filters are shown in the Appendix~\ref{sec:calibrations_gr}. 

We observe that the strength of the corrections and their positions are compatible with the findings we discussed previously in the validation section. In that section we noted that in any range of shape and S\'ersic index the uncalibrated measurements of the sub-populations of galaxies at the faintest magnitude range present overestimated half light radii and S\'ersic Indices. In the calibration map they are assigned with larger vector corrections in size, which calibrate the measurements towards smaller values. If the correction in size is small, then we observe that a calibration in S\'ersic Index is applied, where the recovered value was larger than the model parameter. The same observations are valid also for the other two filters (shown in Appendix~\ref{sec:calibrations_gr}). 
The fact that the measurements and their associated corrections are similar across photometric bands indicates that our final set of calibrated results are robust to the survey characteristics, such as overall PSF size and noise level, that vary between bands. Furthermore, the vast majority of cells across all three calibration maps show little corrections, suggesting that our converged fits are in general reliable and represent the light profiles well. 

\section{Non Parametric fits}
\label{sec:nonparams}

\subsection{ZEST+ Setup}
\label{sec:zestp}
\textsc{ZEST+} is a C++ software application which uses a non-parametric approach to quantify galaxy structure and perform morphological classification. It is based on the \textsc{ZEST} algorithm by \citealt{Scarlata2006,Scarlata2007}, which saw a first application in \citealt{Cameron}. Compared with its predecessor, \textsc{ZEST+} has increased execution speed.
The software architecture consists of two main modules: \emph{Preprocessing} and \emph{Characterization}. The former performs image cleaning, main object centring and segmentation, the latter calculates structure and substructure morphological coefficients.

\subsubsection{Preprocessing}
\label{sec:zest_pre}
In this module the algorithm uses the stamps and the input catalogue provided by the SAND routine. The input catalogue includes the coordinates and the geometrical parameters of the target galaxy and its neighbours in order to remove nearby objects, subtract the background, determine the centre of the galaxy and measure its Petrosian radius. \\
The Petrosian radius is defined as the location where the ratio of flux intensity at that radius, $I(R)$, to the mean intensity within the radius, $\langle I(<R)\rangle$, reaches some value, denoted by $\eta(R)$ \citep{Petrosian}:
\begin{equation}
\mathcal{\eta = \frac{I(R)}{\langle I(R) \rangle}}.
\label{eq:petrosian}
\end{equation}
For this work the Petrosian radius corresponds to the location where $\eta(R)=0.2$. The Petrosian ellipse associated with the object contains the pixels which are used in the \emph{Characterization} module to calculate the morphological coefficients of the central galaxy. 

\subsubsection{Characterization}
The measurements provided by ZEST+ are galaxy concentration (C), asymmetry (A), clumpiness or smoothness (S) and Gini (G) and $M_{20}$ coefficients. This set of parameters, which we refer to as to the \textit{CASGM system}, quantifies the galaxy light distribution and is widely used in studies which correlate the galaxy structure to other parameters, such as colour and peculiar features indicating mergers or galaxy interactions (see for example \citealt{Conselice2000}, \citealt{Conselice2003}, \citealt{Lotz2004} and \citealt{Zamojski2007}); other similar quantities have been recently introduced by \citet{Freeman}. \\
\\
The \textit{concentration} of light, first introduced in \citealt{Bershady2000} and \citealt{Conselice2003}, expresses how much light is in the centre of a galaxy as opposed to its outer parts; it is defined as
\begin{equation}
C = 5\log\left(\frac{r_{80}}{r_{20}}\right),
\label{eq:concentration}
\end{equation}
where $r_{80}$ and $r_{20}$ are the elliptical radii enclosing, respectively, the $20\%$ and $80\%$ of the flux contained within the Petrosian ellipse of the object. ZEST+ outputs three different values of concentration, $C$, $C_{ext}$ and $C_{circ}$. The first parameter is calculated using the total flux measured within the Petrosian ellipse, the second using the flux given as input by the user within the same ellipse and the third one using the Petrosian flux within a circular aperture. For this work we refer to $C$ as the concentration. \\
\\
The \textit{asymmetry} is an indicator of what fraction of the light in a galaxy is in non-asymmetric components. Introduced in \citealt{Schade1995} first, and then in \citealt{Abraham} and \citealt{Conselice1997} independently, asymmetry is determined by rotating individual galaxy images by $180^\circ$ about their centres and self-subtracting these from the original galaxy images. This procedure is applied after the \textit{Preprocessing} module, where the background is $\kappa \sigma-$clipped and subtracted. The value of pixel $(i,j)$ in the subtracted image is calculated as:
\begin{equation}
\Delta I(i,j) = I(i,j) - I_{180}(i,j) = I(i,j) - I(2i_c-i, 2j_c -j),
\end{equation}
where $I_{180}$ is the rotated image and $(i_c, j_c)$ are coordinates of the centre of the galaxy. \\
To take into account the asymmetry of the background, ZEST+ works with smoothed images of the galaxies and their rotated version. In this method, proposed in \citealt{Zamojski2007}, the smoothed image is obtained through a five-point convolution:
\begin{equation}
f_{i,j}^{S} = \frac{1}{5} (f_{i,j}+f_{i+1,j}+f_{i-1,j}+f_{i,j+1}+f_{i,j-1}),
\label{eq:flux_smoothed_image}
\end{equation}
where $f_{i,j}$ is the flux at the $(i,j)$ pixel of the image, and $f_{i,j}^{S}$ is the flux in the same coordinates after the smoothing.
The asymmetry of the original image is defined as
\begin{equation}
A_0 = \frac{1}{2} \frac{\sum_{i,j}|I(i,j)-I_{180}(i,j)|}{\sum_{i,j}|I(i,j)|},
\label{eq:asymmetry0}
\end{equation}
where $I(i,j)$ and $I_{180}(i,j)$ express the intensity of the flux at the pixel (\textit{i,j}) in the original and rotated image, respectively. Similarly we define the asymmetry of the smoothed image:
\begin{equation}
A_{0,S} = \frac{1}{2} \frac{\sum_{i,j}|I^{S}(i,j)-I^{S}_{180}(i,j)|}{\sum_{i,j}|I^{S}(i,j)|}.
\label{eq:asymmetry1}
\end{equation}
Assuming that the intrinsic asymmetry of the light does not change in the smoothed version, we consider that the difference between the two values of asymmetry is due to the background. Smoothing reduces the standard deviation of the background by a factor $\sqrt{5}$ with respect to its un-smoothed version. The combination of $A_0$ and $A_{0,S}$ then gives the final asymmetry value:
\begin{equation}
\label{eq:FAsymmetry}
A=A_0-\frac{A_0-A_{0,S}}{1-1/\sqrt{5}},
\end{equation}
where the subtracted term corresponds to the background correction factor.\\
\\
The \textit{clumpiness} or \textit{smoothness} parameter, introduced in \citealt{Conselice2003}, describes the fraction of light which is contained in clumpy distributions. Clumpy galaxies show a large amount of light at high spatial frequencies, and smooth systems at low frequencies. This parameter is therefore useful to catch patches in the galaxy light which reveal star-forming regions and other fine structure. ZEST+ calculates the clumpiness by subtracting a smoothed image, $I_{S}(i,j)$, from the original, $I(i,j)$, and then quantifying the residual image, $I_{\Delta}(i,j)$. The smoothed image is obtained by convolving the original image with a Gaussian filter of FWHM equal to 0.25 times the Petrosian radius calculated during the \emph{Preprocessing} module. In $I_{\Delta}(i,j)$ the clumpy regions are quanitifed from the pixels with intensity higher than $k=2.5$ times the background standard deviation in the residual image $\sigma_\Delta$. These pixels are then used to calculate the clumpiness of the galaxy:
\begin{equation}
\label{eq:Clumpiness}
S= \frac{\sum_{i,j}I_{\Delta}(i,j)}{\sum_{i,j}|I(i,j)|}_{I_{\Delta}(i,j)>k\sigma_{\Delta}} .
\end{equation}
\\
Similarly, the \textit{Gini} coefficient quantifies how uniformly the flux of an object is distributed among its pixels. A Gini coefficient $G=1$ indicates that all the light is in one pixel, while $G=0$ means that every pixel has an equal share. To calculate \textit{Gini} ZEST+ uses the definition by \citet{Lotz2004,Lotz2008P2,Lotz2008P1}: 
\begin{equation}
\label{eq:Gini}
G = \frac{1}{\hat{I}n(n-1)}\sum^n_i(2i-n-1)\hat{I}_i,
\end{equation}
where $\hat{I}$ is the mean flux of the galaxy pixels and $\hat{I}_i,$ indicates the flux in the $ith$ pixel, sorted by increasing order. \\
\\
The $M_{20}$ coefficient is similar to the concentration \textit{C} in that its value indicates the degree to which light is concentrated in an image; however a high light concentration (denoted by a very negative value of $M_{20}$) doesn't imply a central light concentration. For this reason it is useful in describing the spatial distribution of bright substructures within the galaxy, such as spiral arms, bars or bright nuclei. The computation of this parameter requires first that the pixels within the Petrosian ellipse of the galaxy are ordered by flux; then the $20\%$ brightest pixels are selected and for each pixel \textit{i} the second-order moments are calculated:
\begin{equation}
E_{i} = I_{i}[(x_i - x_c)^{2} + (y_i - y_c)^{2}],
\end{equation}
where $I_i$ is the flux in the $i-th$ pixel, $(x_i,y_i)$ the coordinates of the pixel and $(x_c,y_c)$ the coordinates of the centre of the Petrosian ellipse.
The sum of these moments is $E=\sum^{N_{20}}_{i}E_i$, where $N_{20}$ is the multiplicity of the  $20\%$ brightest selected pixels. Given $E_{tot}$ as the sum of the second order moments of all the pixels in the ellipse, we finally calculate $M_{20}$ as:
\begin{equation}
\label{eq:M20}
M_{20} = log \frac{E}{E_{tot}}.
\end{equation}

\begin{figure}
	\includegraphics[width=\columnwidth]{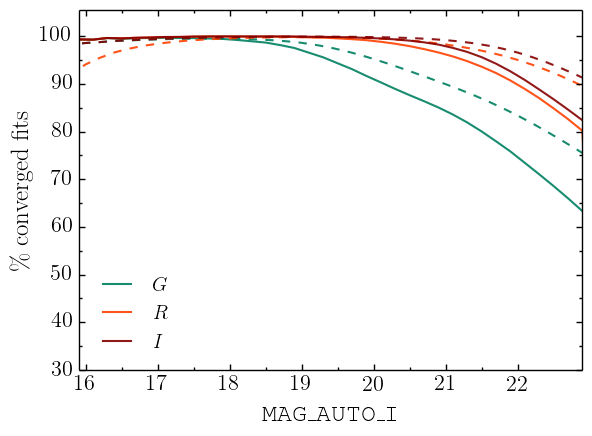}
	\caption{Fitting completeness of non-parametric converged fits in the \textit{g}, \textit{r} and \textit{i} bands, expressed in terms of the percentage of converged fits in bins of 0.2 magnitude, normalised on the total number of selected objects in that magnitude bin. By \textit{converged fits} we refer in this case to the objects flagged by \textsc{ZEST+} as fits without errors, either during the cleaning process or the characterization routine, as described in more detail in Section~\ref{subsec:npcompl}. Magnitude-limited completeness is represented by the dashed lines. We obtain almost full recovery in the \textit{i} and \textit{r} filters up to $i\sim21$, losing only a few saturated objects.}
	\label{fig:completeness_zest}
\end{figure}

\subsection{Completeness}
\label{subsec:npcompl}
The measurements of Gini, M20, Concentration, Asymmetry and Clumpiness are matched with diagnostic flags which inform the user whether errors occurred during the cleaning step of the process or in the calculation of the coefficients. To be more precise, the flag \texttt{Error} (we label it in our catalogue as \texttt{ERRORFLAG}) indicates whether a problem occurred while processing an object: if it is non-zero, it traces an error encountered during the calculation of the structural parameters, and flags the measurements as not reliable. The \textit{contamination flag} informs the user whether the cleaning process was unsuccessful due to the presence of a neighbour covering the centre of the galaxy; in this case the program outputs $contamination flag =-2$. Therefore in this test we considered as converged fits the measurements with $\texttt{ERRORFLAG}=0 \wedge contamination \ flag \neq -2$. Then we define the fitting completeness as we did for the parametric fits, following Equation~\ref{eq:compl_eq}. \\

The results for the \textit{g}, \textit{r} and \textit{i} bands are shown in Figure~\ref{fig:completeness_zest}. With the cut in \texttt{ERRORFLAG} and \textit{contamination flag} we discard a total of $\sim10\%$ of objects. We observe some fluctuations at the brightest end, where we find cases of large bright galaxies whose Petrosian ellipses were underestimated or cases with saturated objects, and at the faintest end, where it is more common to have higher noise contamination within the Petrosian ellipse. The overall number of successful fits is more than $\sim90\%$ in the \textit{i} and \textit{r} filters and $\sim80\%$ in the \textit{g} band. The dashed lines show magnitude-limited, rather than differential, completeness. \\

\begin{figure*}
	\includegraphics[width=\textwidth]{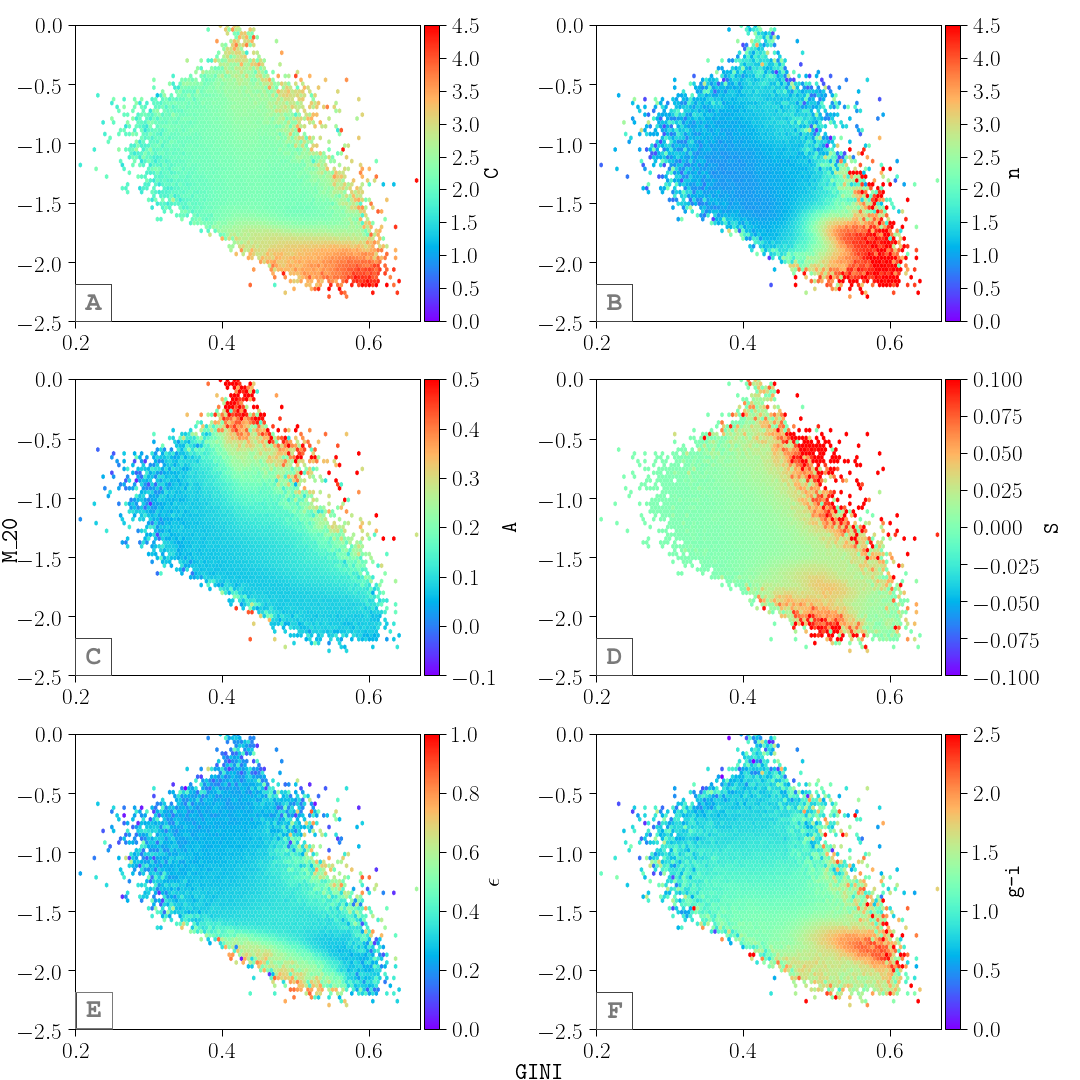}
	\caption{Gini-M20 relation shown as a function of Concentration C (Panel A), S\'ersic Index (Panel B), Asymmetry A (Panel C), Clumpiness S (Panel D), ellipticity $\epsilon$ (Panel E) and $g-i$ colour (Panel F). The expected trends for the relations and their gradients are recovered, as discussed in more detail in Section~\ref{subsec:npvalidation_scv}.}
	\label{fig:ginim20}
\end{figure*}

\subsection{Validation}
\label{subsec:npvalidation_scv}
By way of a simple internal validation, we show in Figure~\ref{fig:ginim20} the uncalibrated measurements from \textsc{ZEST+} and the relationships between them (only the Concentration parameter is calibrated). In particular we focus on the Gini-M20 relation, studied as a function of other morphological parameters: Concentration (C), Clumpiness (S) and Asymmetry (A), shown in Panels A, C and D, respectively. Since we can benefit from the additional information provided by parametric fitting, we show the same relation as a function of calibrated parametric quantities: S\'ersic Index $n$ (Panel B), ellipticity $\epsilon$ (Panel E) and $g-i$ colour (Panel F).

In the cross-comparison between non-parametric measurements, we observe that even though those are still un-calibrated, we can easily recover the expected trends with very few outliers. As an example consider the first panel, where the Gini-M20 relation is colour-coded by the Concentration. The objects with low M20 values present a high concentration of light; from the figure we observe that in the Gini-M20 plane these objects tend to have larger values of Gini, which means that the light is not uniformly distributed. If we now consider the third parameter, we notice that the Concentration (and the S\'ersic Indexes) of these objects lies in its highest range: this explains that the light of these galaxies is very concentrated, and located at the centre of the galaxy.

From panels C, D and E we add the expected information that these objects are also symmetric, lack clumpy regions and are mostly rounded. These observations were further confirmed by our visual inspection of image stamps. If the combined analysis of the first five panels helps us to distinguish between two different morphological regions in the Gini-M20 plane, Panel F shows a colour bi-modality which overlaps with the morphological one: disk-like galaxies tend to be bluer and the bulge-dominated ones are redder. Finally, we perform a qualitative comparison with the CAS-GM measurements made by \cite{Zamojski2007} using high-resolution Hubble Space Telescope data (their Figures 3 and 17). The range of values for Gini and M20 are much the same for the bulk of the population, though our far larger sample explores more extreme values of low Gini coefficient and less negative M20. The correlation between M20 and asymmetry, at ${rm M}\_20>-2$, is also clearly present in Figure~~\ref{fig:ginim20}, panel C. We expect the PSF to suppress fine substructure, and the trend between clumpiness and Gini coefficient in our sample is not as clear as that found by \cite{Zamojski2007}. Nevertheless, redder galaxies do tend to avoid regions of high clumpiness, as expected. 

\begin{figure*}
	\includegraphics[width=\textwidth]{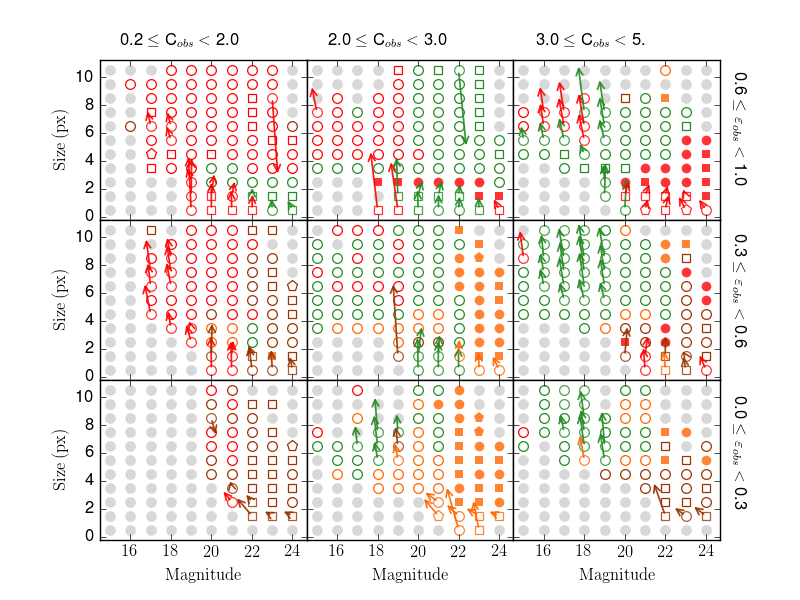}
	\caption{Calibration map for the non-parametric measurements in the \textit{i} band, obtained through the simulation routine described in Section~\ref{subsec:npsimulations}. The calibrations are determined in a 4D parameter space, where the correlation of size, magnitude, ellipticity and Concentration between the measured values and the model parameters is studied. The information in the map is displayed using different symbols and colours with the same \textsc{Galfit} adopted for the parametric fits.  They calibrations are presented in a size-magnitude plane, divided in different cells according to the shown sub-ranges in ellipticity and Concentration. The components of the correction vectors are the magnitude discrepancy $\eta^{mag}$ on the x axis and the size discrepancy $\eta^{size}$ on the y axis, according to the definitions given in Equations~\ref{eq:calsmodel} and~\ref{eq:calsfits}. If these corrections are small ($\eta^{mag} < 0.1  \wedge \  \eta^{size}<10 \%$) the length of the arrow is set to zero and only a symbol identifies them. If the scatter in ellipticity ($\epsilon$) or Concentration ($C$)  is large ($\eta^{\epsilon}>0.1$ and $\eta^{C} > 20\%$, respectively), then the symbol is coloured in orange or red, respectively. If the calibration is large in both parameters, it is coloured in brown. The symbol is empty if the ZEST+ recovered value is smaller than the model. Different shapes are used referring to the total scatter (w) in the 4D parameter space of the model parameters, defined in Equation~\ref{eq:variance}; the symbol is a pentagon if $w>1.5$ and a square if $w>1$, otherwise it is a circle.}
	\label{fig:npmapcalsnew}
\end{figure*}

\subsection{Calibrations and diagnostics of the corrected results}
\label{subsec:npsimulations}
In order to apply corrections to the non-parametric measurements, which are crucial in accounting for the impact of the PSF, we adopt the same approach used for the parametric fits: we consider the images from the \textit{UFig-BCC} release for DES Y1 and treat them as if they were real data, as explained in detail in Section~\ref{sec:simulations}. We then derive calibration maps exactly as described in Section~\ref{subsec:correction_calc}, determining the correction for each parameter of interest as the discrepancy between the central values of the model and the fitting results distributions in each cell. The equations~\ref{eq:calsmodel},~\ref{eq:calsfits},~\ref{eq:cals} and~\ref{eq:variance} are valid also in this context, with the exception that the S\'ersic Index, \textit{n}, is now substituted by the Concentration of light, \textit{C}.

In order to derive correction vectors, we first compute \textsc{ZEST+} output parameters for the simulated galaxies {\it before} noise and PSF convolution are applied. We use \textsc{Galfit} to produce noise and PSF-free image stamps based on the \textsc{UFig} model parameters and run \textsc{ZEST+} on them. In this way we construct the truth table of values with which to derive calibration vectors.

Figure~\ref{fig:npmapcalsnew} shows the correction map for the \textit{i} band; the other two filters, \textit{g} and \textit{r}, are presented in Appendix~\ref{sec:calibrations_gr}. Also for non-parametric fits we adopt the same convention of colours and shapes as in Figure~\ref{fig:mapcalsnew}. The length of the arrows is a visual representation of the strength of the vector correction: their x and y components are the discrepancies between the central values of the model distribution and the fitted dataset in each 4-dimensional cell, projected on the size-magnitude plane. When the correction is small ($\eta^{mag} < 0.1  \wedge \  \eta^{size}<10 \%$) a symbol in place of the arrow is shown. Apart from the grey circles, which indicate areas with poor statistics, the colour legend reflects the size of the calibration applied to ellipticity and Concentration. If the scatter in ellipticity or Concentration is large ($\eta^{\epsilon}>0.1$ or $\eta^{C} > 20\%$), then the symbol is coloured in orange or red, respectively. If this condition applies to both parameters simultaneously, it is coloured in brown. If the recovered value underestimates the model input, the symbol is empty. Different shapes are used according to the dispersion $w$ of the 4-dimensional parameter space, calculated considering its covariance matrix, as expressed in Equation~\ref{eq:variance}. Symbols are pentagons when $w>1.5$, squares if $w>1$ and circles otherwise. 

We observe that the majority of red cells, where a larger correction in Concentration is required, have an empty symbol: this tells us that ZEST+ tends to recover underestimated values of concentration. This behaviour is entirely expected, due to the fact that \textsc{ZEST+} cannot account the PSF in computing results. We demonstrate this aspect more explicitly in Figure~\ref{fig:nCnew}, which shows the relation between the S\'ersic Index and the Concentration before (grey contours) and after (magenta) applying the corrections. For clarity, we have removed objects where the pixel size significantly hampers our ability to measure the concentration (i.e. where $\texttt{FLUX\_RADIUS} < 2.5~$px). The solid blue line in this figure is the analytic relationship between S\'ersic index and concentration, adapted from \cite{Graham05} for the case of measurements within the Petrosian radius. The \textit{flattening} effect we observe in the uncalibrated population of Concentration values reflects exactly what we observe in the calibration map and through the corrections we obtain values that are much more consistent with expectations. This test shows that using calibrated values from both parametric and non-parametric approaches to quantifying galaxy structure allows us to use the advantages of both methods and provide a firmer grip on the characteristics of the galaxy population. We will exploit the strength of our dual-method, multi-band morphology catalogue in a series of future papers.

\begin{figure}
	\includegraphics[width=\columnwidth]{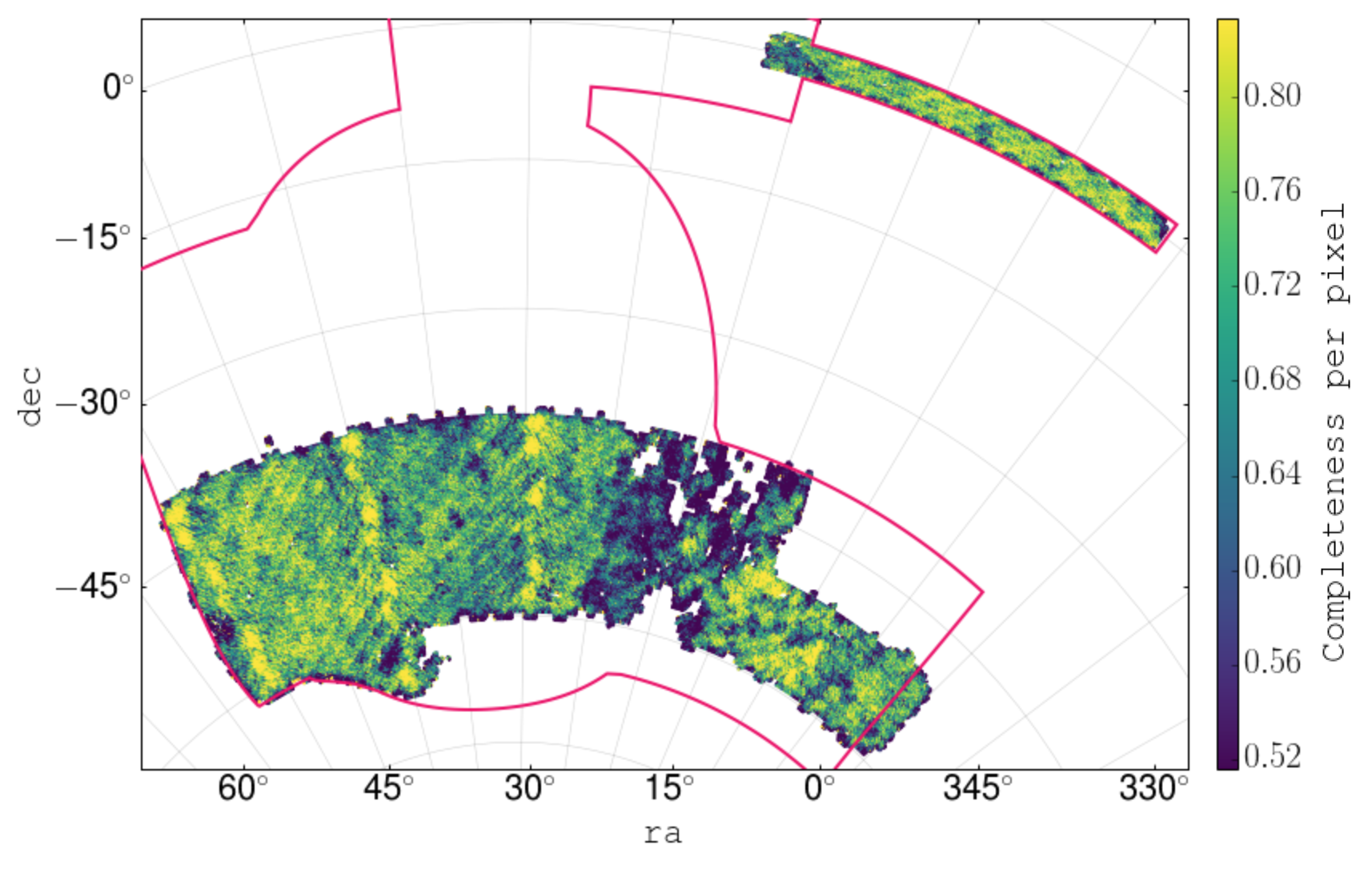}
	\caption{Healpix map of the ratio between two galaxy samples. We apply to the \texttt{Y1A1} data the sample-selection cuts to obtain the first sample, and then apply the science-ready cuts to it in order to get the second one. The ratio gives the completeness per pixel of the science-ready sample.}
	\label{fig:finalmap}
\end{figure}

\begin{figure}
	\includegraphics[width=\columnwidth]{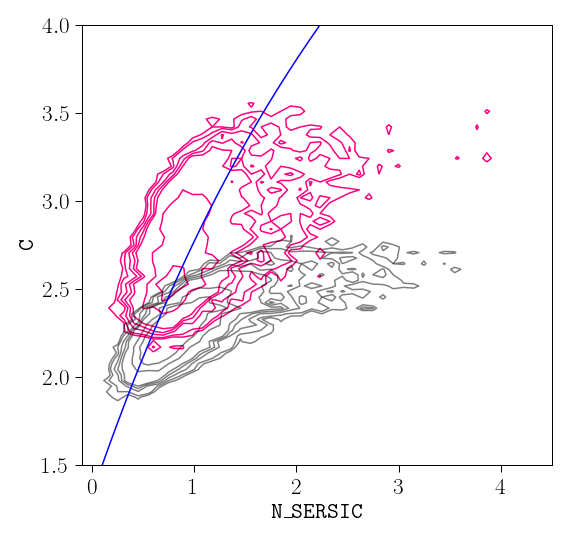}
	\caption{S\'ersic Index-Concentration relation before (grey) and after (magenta) applying the calibrations. The solid blue line show the analytic relationship between S\'ersic index and concentration. The \textit{flattening effect} present in the un-calibrated measurements is due to PSF effects which is corrected by our calibration.}
	\label{fig:nCnew}
\end{figure}

\section{Science-ready cuts}

We finish by summarising the overall selection function of the galaxy sample and detail a set of simple cuts that could form the basis of a sample for scientific analysis. We exclude from consideration objects that meet any one of the following criteria:
\begin{itemize}
\item $\texttt{SExtractor FLAGS} > 0$
\item $\texttt{CLASS\_STAR} > 0.9$
\item $\texttt{MAG\_AUTO\_I} > 23$
\item $\texttt{FLUX\_RADIUS} \le 0$
\item $\texttt{KRON\_RADIUS} \le 0$
\item $\texttt{FLAGS\_BADREGION} > 0$
\item Objects with a neighbour that overlaps $50\%$ or more of its expanded Kron ellipse. The relevant column in the catalogue for this criterion is \texttt{MAX\_OVERLAP\_PERC}.
\item Objects that have unrecoverable errors in the \textsc{SExtractor} output of their neighbouring objects (if any). 
\end{itemize}
This initial sample comprises 45 million objects over $1800$ square degrees that is $80\%$ complete in S\'ersic measurements up to magnitude 21.5.

\noindent To prepare a high completeness science-ready galaxy sample, we suggest the following initial cuts. Science problems requiring higher completeness and/or greater uniformity across the footprint will require additional cuts, dependent on the goals. In some circumstances fainter galaxies could also be included in the sample.
\begin{itemize}
\item $\texttt{MAG\_AUTO\_I} \le 21.5$
\item $S/N > 30$
\item $\texttt{SPREAD\_MODEL} + 1.67 \times \texttt{SPREADERR\_MODEL} > 0.005$
\end{itemize}
For the i-band catalogue, these cuts produce a sample of 12 million galaxies that is $90\%$ complete in S\'ersic measurements and $99\%$ complete in non-parametric measurements. \\
In Fig.~\ref{fig:finalmap} we show a ratio of two healpix maps realised with two samples. We first applied the cuts used for the sample selection, with an additional cut in $\texttt{MAG\_AUTO}<21.5$. We chose this threshold according to the analysis of the completeness discussed in Section~\ref{sec:parametric_completeness}. Then we select from this sample all the objects with pass the set of science-ready cuts we proposed above. 
The map shows the completeness per pixel, which is overall uniform. It also guides the catalogue users to possibly select specific areas for future analyses.

\section{Conclusions}
\label{sec:conclusions}
We have presented the process of preparing, producing and assembling the largest structural and morphological galaxy catalogue to date, comprising $45$ million objects over $1800$ square degrees, which are taken from the first year of the Dark Energy Survey observations (DES Y1). We adopted both parametric and non-parametric approaches, using \textsc{Galfit} and \textsc{ZEST+}.
In order to optimize their performance according to the characteristics of our sample, in particular in those cases where the galaxy we want to fit has one or more close neighbours, we developed a neighbour-classifier algorithm as part of a pre-fitting pipeline (Section~\ref{sec:prefittingpip}) which automatically prepares the postage stamps and all the settings required to simultaneously fit the objects in the presence of overlapping isophotes. We stress the importance of this step because a precise treatment of the size of the stamps and the neighbouring objects allows the recovery of more accurate measurements.\\
In Section~\ref{sec:parametric_completeness} we presented the fitting completeness of the parametric fits in the \textit{g}, \textit{r} and \textit{i} filters as a function of object magnitude. Using a tile-by-tile analysis, we show that the highest percentages of non-converged fits are localised at the West and East borders of the footprint, where there is a high stellar density due to the vicinity of the Large Magellanic Cloud. After applying star-galaxy separation based on a linear combination of the parameter \texttt{SPREAD\_MODEL} and its uncertainty, we find that the fitting efficiency remains high ($>80\%$) up to magnitude $<22$ for the \textit{i} and \textit{r} band, and magnitude $<21$ for the \textit{g} band. We also studied the subsequent fitting completeness in relation to survey data characteristics that are expected to impact the performance of \textsc{Galfit}: stellar density, PSF FWHM and image depth. We conclude that at relatively bright magnitudes ($i<21.5$) the completeness has a relatively weak dependence on these quantities, and high completeness can be maintained without much loss of survey area.\\
In Section~\ref{sec:validation} we analysed the properties of the converged fits, isolating a small fraction ($<5\%$) of outliers in magnitude recovery, and a branch of objects with high S\'ersic indices and large radii that we believe to be spurious. Removing low S/N galaxies efficiently cleans the sample of these populations. Following this basic validation, we calibrate the S\'ersic measurements using state-of-the-art \textsc{UFig} image simulations, deriving correction vectors via the comparison of input model parameters and the resulting fits by \textsc{Galfit}. In Section~\ref{sec:nonparams} we repeated the above mentioned diagnostics for the non-parametric fits, benefiting from the internal diagnostic flags provided by ZEST+ itself in order to quantify the quality of the image and so the reliability of the measurements.
For the non-parametric dataset we adopted the same method to derive the calibrations described in Section~\ref{sec:simulations}, finding that corrections are stronger for low signal to noise galaxies, similar to the parametric case. In particular, we highlight the calibration of galaxy concentration, which is adversely affected due to fact that ZEST+ cannot account for the PSF.\\
Finally, we summarised the selection function and a recommended set of cuts to form a basic science sample. Our catalogue represents a valuable instrument to explore the properties and the evolutionary paths of galaxies in the DES Y1 survey volume, which will be used in a series of forthcoming publications.

\section*{Acknowledgements}
FT would like to thank Sandro Tacchella for his useful suggestions about fitting calibration and fruitful discussions.\\

Funding for the DES Projects has been provided by the U.S. Department of Energy, the U.S. National Science Foundation, the Ministry of Science and Education of Spain, 
the Science and Technology Facilities Council of the United Kingdom, the Higher Education Funding Council for England, the National Center for Supercomputing 
Applications at the University of Illinois at Urbana-Champaign, the Kavli Institute of Cosmological Physics at the University of Chicago, 
the Center for Cosmology and Astro-Particle Physics at the Ohio State University,
the Mitchell Institute for Fundamental Physics and Astronomy at Texas A\&M University, Financiadora de Estudos e Projetos, 
Funda{\c c}{\~a}o Carlos Chagas Filho de Amparo {\`a} Pesquisa do Estado do Rio de Janeiro, Conselho Nacional de Desenvolvimento Cient{\'i}fico e Tecnol{\'o}gico and 
the Minist{\'e}rio da Ci{\^e}ncia, Tecnologia e Inova{\c c}{\~a}o, the Deutsche Forschungsgemeinschaft and the Collaborating Institutions in the Dark Energy Survey. 

The Collaborating Institutions are Argonne National Laboratory, the University of California at Santa Cruz, the University of Cambridge, Centro de Investigaciones Energ{\'e}ticas, 
Medioambientales y Tecnol{\'o}gicas-Madrid, the University of Chicago, University College London, the DES-Brazil Consortium, the University of Edinburgh, 
the Eidgen{\"o}ssische Technische Hochschule (ETH) Z{\"u}rich, 
Fermi National Accelerator Laboratory, the University of Illinois at Urbana-Champaign, the Institut de Ci{\`e}ncies de l'Espai (IEEC/CSIC), 
the Institut de F{\'i}sica d'Altes Energies, Lawrence Berkeley National Laboratory, the Ludwig-Maximilians Universit{\"a}t M{\"u}nchen and the associated Excellence Cluster Universe, 
the University of Michigan, the National Optical Astronomy Observatory, the University of Nottingham, The Ohio State University, the University of Pennsylvania, the University of Portsmouth, 
SLAC National Accelerator Laboratory, Stanford University, the University of Sussex, Texas A\&M University, and the OzDES Membership Consortium.

Based in part on observations at Cerro Tololo Inter-American Observatory, National Optical Astronomy Observatory, which is operated by the Association of 
Universities for Research in Astronomy (AURA) under a cooperative agreement with the National Science Foundation.

The DES data management system is supported by the National Science Foundation under Grant Numbers AST-1138766 and AST-1536171.
The DES participants from Spanish institutions are partially supported by MINECO under grants AYA2015-71825, ESP2015-66861, FPA2015-68048, SEV-2016-0588, SEV-2016-0597, and MDM-2015-0509, 
some of which include ERDF funds from the European Union. IFAE is partially funded by the CERCA program of the Generalitat de Catalunya.
Research leading to these results has received funding from the European Research
Council under the European Union's Seventh Framework Program (FP7/2007-2013) including ERC grant agreements 240672, 291329, and 306478.
We  acknowledge support from the Australian Research Council Centre of Excellence for All-sky Astrophysics (CAASTRO), through project number CE110001020, and the Brazilian Instituto Nacional de Ci\^encia
e Tecnologia (INCT) e-Universe (CNPq grant 465376/2014-2).

This manuscript has been authored by Fermi Research Alliance, LLC under Contract No. DE-AC02-07CH11359 with the U.S. Department of Energy, Office of Science, Office of High Energy Physics. The United States Government retains and the publisher, by accepting the article for publication, acknowledges that the United States Government retains a non-exclusive, paid-up, irrevocable, world-wide license to publish or reproduce the published form of this manuscript, or allow others to do so, for United States Government purposes.




\bibliographystyle{mnras}
\bibliography{biblio}

\appendix
\section{Calibration maps for the \textit{g} and \textit{r} filters}
\label{sec:calibrations_gr}
In this Appendix we present the calibration maps for both parametric and non-parametric measurements in the \textit{g} and \textit{r} bands. They were obtained following the procedure described in Sections~\ref{sec:simulations}  and~\ref{subsec:npsimulations} for parametric and non-parametric fits, respectively. The maps are displayed following the same conventions adopted for visualising the calibration maps in the \textit{i} band. Those maps are shown in  Figures~\ref{fig:mapcalsnew} and~\ref{fig:npmapcalsnew}.

\onecolumn
\begin{figure*}
	\label{fig:gmapcals}
	\includegraphics[scale=0.77]{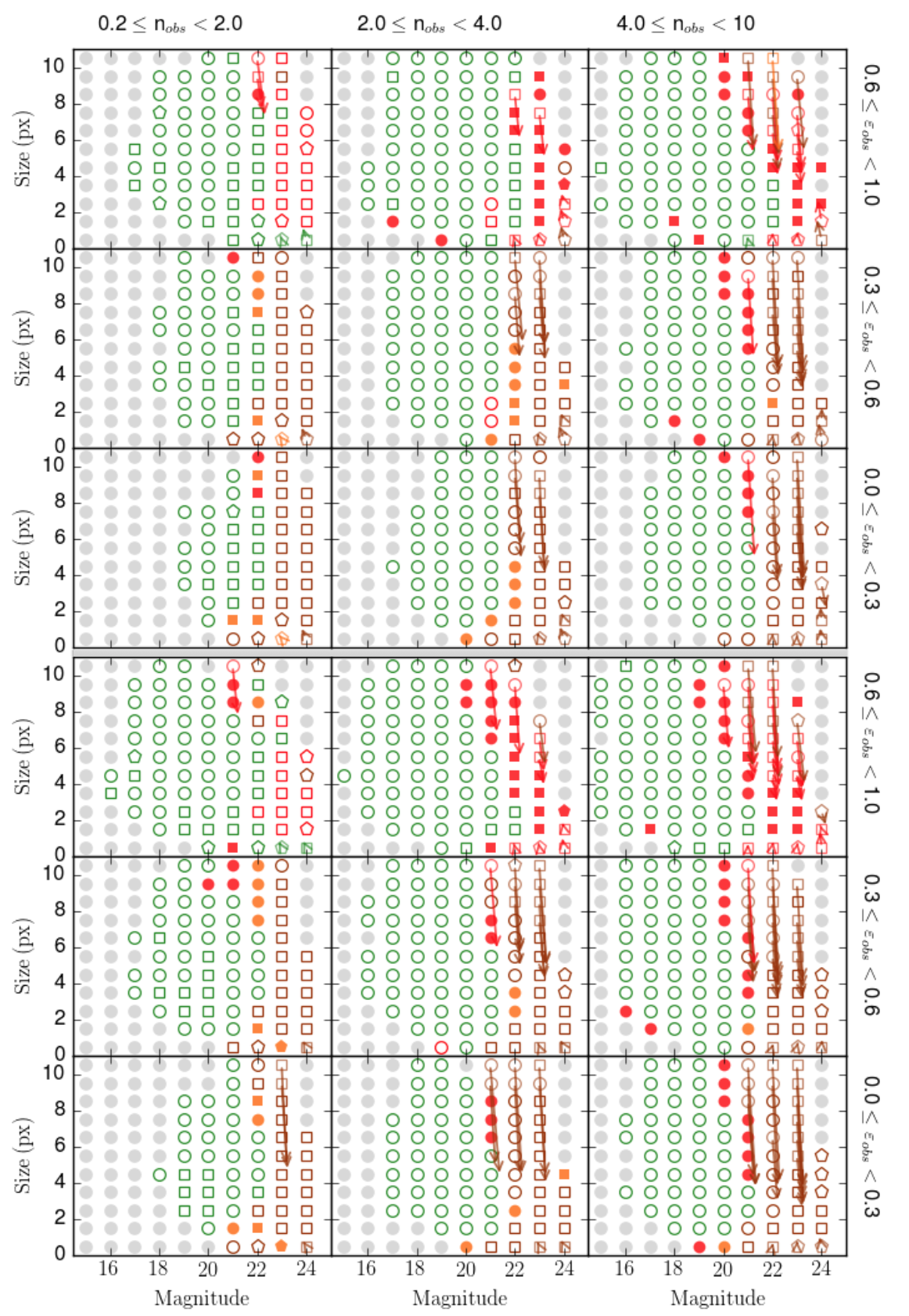}
	\caption{Map of the corrections for S\'ersic parameters in the \textit{g} (upper panel) and \textit{r} (lower panel) filters, obtained through the simulation routine described in Section~\ref{sec:simulations}. Symbols and colours have the same meaning as Figure~\ref{fig:mapcalsnew}.}
	\label{fig:noisemap}
\end{figure*}

\begin{figure*}
	\label{fig:gmapcals}
	\includegraphics[scale=0.77]{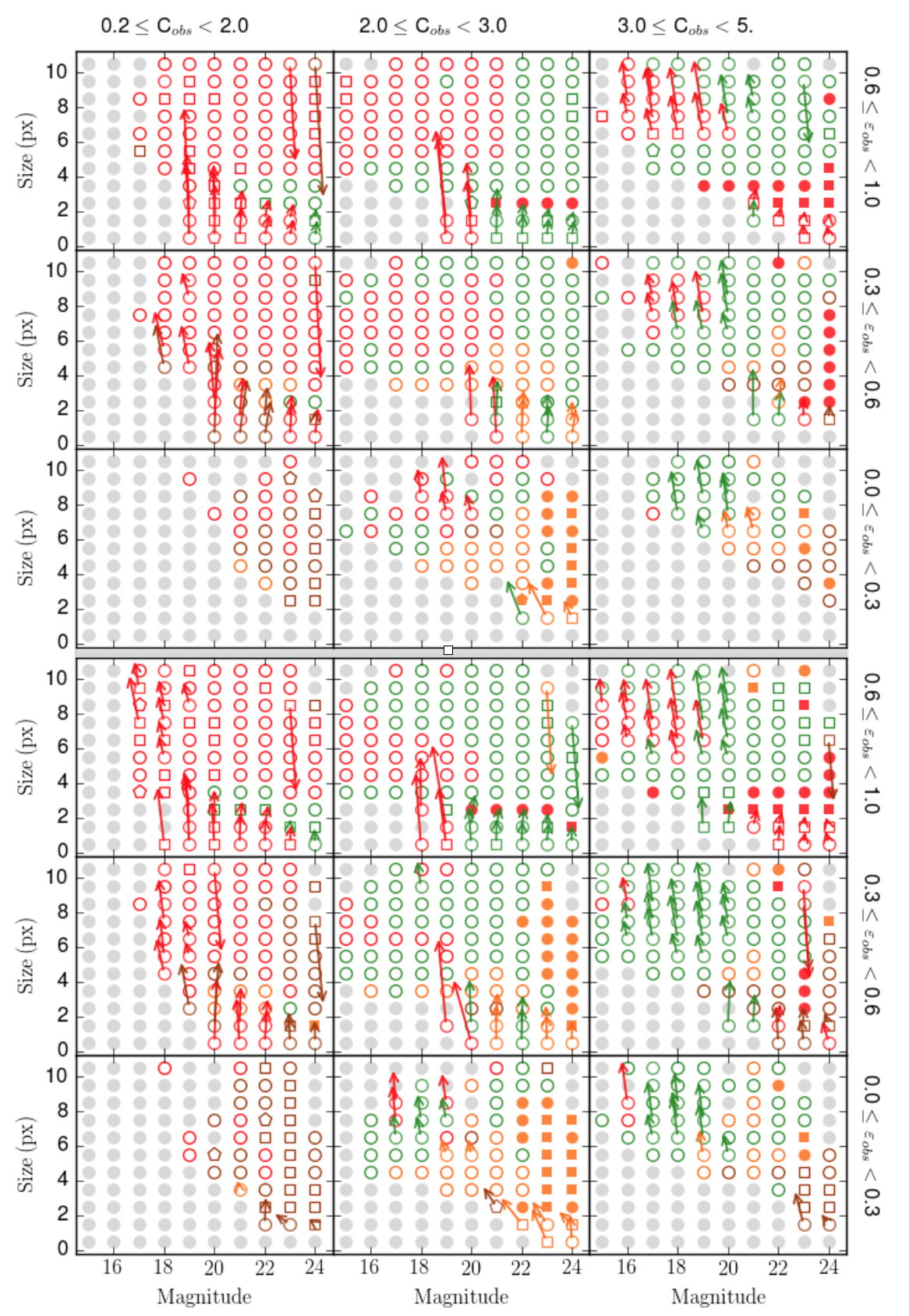}
	\caption{Map of the corrections for ZEST+ output in the \textit{g} (upper panel) and \textit{r} (lower panel) filters, obtained through the simulation routine described in Section~\ref{subsec:npsimulations}. Symbols and colours have the same meaning as Figure~\ref{fig:npmapcalsnew}.}
	\label{fig:noisemap}
\end{figure*}
	
	\onecolumn
	
	\section{Catalog Manual}
	\label{sec:cataloguemanual}
	A description of the columns of the catalogue follows, both for parametric and non-parametric fits. In order to distinguish between filters, the parameters can be labelled with $\_X$, where $X=g, r, i$.
	
	\subsection{Identification columns}
	$\texttt{COADD\_OBJECT\_ID \ - \ }$ Identifier assigned to each object in the co-add DES Y1 dataset, reported here from the \textit{Gold Catalogue}.\\
	$\texttt{TILENAME \ - \ }$ Column reporting the name of the tile image where the object lies.\\
	$\texttt{ID \ - \ }$ Rows enumerator, running for 1 to the total number of entries in the catalogue.\\
	$\texttt{RA \ - \ }$ Right Ascension from the \texttt{Y1A1 GOLD} catalogue.\\
	$\texttt{DEC \ - \ }$ Declination from the \texttt{Y1A1 GOLD} catalogue.\\
	
	\subsection{\textsc{SExtractor} parameters for star-galaxy separation and signal-to-noise}
	$\texttt{SG \ - \ }$ Linear combination of the star-galaxy classifier  $\texttt{SPREAD\_MODEL}$ and its uncertainty, $\texttt{SPREADERR\_MODEL}$, according to Equation~\ref{eq:sgalsep}. A cut in $\texttt{SG>0.005}$ is recommended.	\\
	$\texttt{SN\_X \ - \ }$ Signal-to-noise expressed as the ratio between $\texttt{FLUX\_AUTO\_X}$ and $\texttt{FLUXERR\_AUTO\_X}$.	
	
	\subsection{Columns for Parametric Fits}
	
	\subsubsection{\textbf{Selection and pre-fitting classification flags}}
	$\texttt{SELECTION\_FLAGS\_X \ - \ }$ If equal to 1, then the relative object has been selected, according to the requirements described in Section~\ref{subsec:sampleselection}. It can assume other numerical values in the following cases:
	\begin{itemize}
		\item if the object passes the selection requirements, but is not included in the intersection between the DESDM catalogues and the \texttt{Y1A1 GOLD} catalogue, then this flag is set to 2;
		\item if the object passes the selection requirements, but it is fainter then $\texttt{GOLD\_MAG\_AUTO\_i} = 23$, then the flag is set to 3;
		\item if the object enters in the previous category, but it has no match with the \texttt{Y1A1 GOLD} catalogue, then the flag is set to 4.
	\end{itemize}
	If the object is not selected because it doesn't pass any of the selection requirements, then the $\texttt{SELECTION\_FLAGS\_X}$ and all the other flags are set to zero.\\
	The catalogue version made available to the users includes all the objects which have been selected at least in one of the three bands \textit{g,r,i}.\\
	\color{white} space \color{black} \\
	$\texttt{C\_FLAGS\_X \ - \ }$ Number of neighbours in the fitted stamp. \\
	$\texttt{MAX\_OVERLAP\_PERC\_X \ - \ }$  Percentage of the central galaxy isophotes overlapping with the closest neighbour. If there are no neighbours or no overlapping neighbours, then it is set to 0. A cut in $\texttt{MAX\_OVERLAP\_PERC\_X < 50}$ is recommended. 
	
	\subsubsection{\textbf{Parametric measurements (\textsc{Galfit})}}
	$\texttt{MAG\_SERSIC\_X  \ - \ }$ \textsc{Galfit} value for the magnitude of the galaxy. The value already includes the calibration listed in the column $\texttt{MAG\_CAL\_X}$. \\
	$\texttt{RE\_SERSIC\_X  \ - \ }$ \textsc{Galfit} measure of the half light radius (or Effective radius) of the galaxy. It is expressed in pixels and is already calibrated. The correction is reported in the column $\texttt{RE\_CAL\_X}$.\\
	$\texttt{N\_SERSIC\_X  \ - \ }$ \textsc{Galfit} output for the S\'ersic Index. The measure is calibrated, and the can find the relative correction in the column $\texttt{N\_SERSIC\_CAL\_X}$.\\
	$\texttt{ELLIPTICITY\_SERSIC\_X  \ - \ }$ Ellipticity of the galaxy, calculated by subtracting from unity the \textsc{Galfit} estimate for the axis-ratio. The value is corrected and the calibration is accessible through the column $\texttt{ELLIPTICITY\_SERSIC\_CAL\_X}$.\\
	$\texttt{OUTLIERS\_X  \ - \ }$ If equal to 1, it labels the objects classified as outliers in the catalogue validation process.\\
	$\texttt{FIT\_STATUS\_X  \ - \ }$ If equal to 1, this flag selects all the objects with a successfully validated and calibrated converged fit. \\
	\textbf{Important note:} by applying the recommended cut $\texttt{FIT\_STATUS\_X}=1$, the user is able to collect the sample of validated and calibrated objects in the \texttt{X} filter. This cut is equivalent to applying all together the cuts which are recommended in terms of sample selection, fitting convergence, bad regions masking, exclusion of outliers and significantly overlapping objects, minimization of stellar contamination. A summarising scheme follows:
	\[ (\texttt{FIT\_STATUS\_X=1}) =
	\begin{cases}
	\texttt{FLAGS\_BADREGION=0}  \\
	\texttt{SG>0.005} \\
	\texttt{SELECTION\_FLAGS\_X=1} \\
	\texttt{FIT\_AVAILABLE\_X=1} \wedge \texttt{WARNING\_FLAGS\_CENTRAL\_X=0} \\
	\texttt{MAX\_OVERLAP\_PERC\_X<50} \\
	\texttt{OUTLIERS\_X=0} \\
	\texttt{PARAMETER\_CAL\_X<99} ,
	\end{cases}
	\]
	where the voice \texttt{PARAMETER\_CAL\_X} can be \texttt{MAG\_CAL\_X} etc. In absence of calibration the correction value is set to 99.\\
	For a cleaner sample the user can associate the cut in \texttt{FIT\_STATUS\_X} to the condition \texttt{SN\_X>30}.
	
	\subsection{Columns for non-parametric coefficients (\textsc{ZEST+})}
	$\texttt{SELECTION\_NP\_X  \ - \ }$ If equal to 1, the object is selected in the \texttt{X} filter, otherwise it is 0.\\
	$\texttt{FIT\_STATUS\_NP\_X  \ - \ }$ If equal to 1, this flag selects all the objects with successfully validated and calibrated measurements.\\
	$\texttt{CONCENTRATION\_X  \ - \ }$ \textsc{ZEST+} measurement for the Concentration of light. See Equation~\ref{eq:concentration} for its definition. The calibration vector is listed in the column $\texttt{CONCENTRATION\_CAL\_X}$.\\
	$\texttt{ASYMMETRY\_X  \ - \ }$ \textsc{ZEST+} value for the Asymmetry (see Equation~\ref{eq:FAsymmetry}). \\
	$\texttt{CLUMPINESS\_X  \ - \ }$ \textsc{ZEST+} value for the Clumpiness (see Equation~\ref{eq:Clumpiness}). \\
	$\texttt{GINI\_X  \ - \ }$ Measure of the Gini parameter, defined in Equation~\ref{eq:Gini}.\\
	$\texttt{M20\_X  \ - \ }$ Measure of the M20 parameter, for more details see Equation~\ref{eq:M20}.\\


\bsp	
\label{lastpage}
\end{document}